\documentclass[12pt]{article}
\pdfoutput=1

\usepackage{draft}
\usepackage{enumitem}

\usepackage{multirow}

\usepackage{graphicx,etoolbox}
\usepackage[all]{xy}
\robustify{\includegraphics}
\usepackage{color}
\usepackage{tcolorbox}
\usepackage{tikz-cd} 
\usepackage[all]{xy}
\usepackage{array,multirow}
 \usepackage{youngtab}
 \usepackage{amssymb,amsthm}
\usepackage[nosort]{cite}
\DeclareMathOperator{\lcm}{lcm}

\makeatletter
\newcommand\xleftrightarrow[2][]{%
  \ext@arrow 9999{\longleftrightarrowfill@}{#1}{#2}}
\newcommand\longleftrightarrowfill@{%
  \arrowfill@\leftarrow\relbar\rightarrow}
\makeatother

\newtheorem{theorem}{Theorem}

\newtheorem{conj}[theorem]{Conjecture}
\theoremstyle{definition}
\newtheorem{eg}[theorem]{Example}

\newcommand{\neswarrow}{%
  \mathrel{\text{\ooalign{$\swarrow$\cr$\nearrow$}}}%
}
\newcommand{\nwsearrow}{%
  \mathrel{\text{\ooalign{$\searrow$\cr$\nwarrow$}}}%
}

\usepackage{blkarray}% http://ctan.org/pkg/blkarray

\include{mymacros}

\usepackage[normalem]{ulem}

\begin{document}

\begin{titlepage}

\preprint{PUPT-2555}

\begin{center}

\hfill \\
\hfill \\
%
%\title{ 
%Twisted irregular codimension-two defects in 6d (2,0) theories, Argyres-Douglas theories and chiral algebras
%}

%\title{  
%Argyres-Douglas Theories  from 
%Outer-Automorphism Twist
%of 6d $(2,0)$ Theories
%}

\title{ 
Codimension-two defects and Argyres-Douglas theories from 
outer-automorphism twist
in 6d $(2,0)$ theories
}

\author{Yifan Wang$^a$ and Dan Xie$^{b,c}$
}
 \address{${}^a$Joseph Henry Laboratories\\ Princeton University, Princeton, NJ 08544, USA}
\address{${}^b$Center of Mathematical Sciences and Applications, Harvard University, Cambridge, 02138, USA}
\address{${}^c$Jefferson Physical Laboratory, Harvard University, Cambridge, MA 02138, USA}

\email{yifanw@princeton.edu, dxie@cmsa.fas.harvard.edu}

\end{center}

\abstract{The 6d $(2,0)$ theory has codimension-one symmetry defects associated to the outer-automorphism group of the underlying ADE Lie algebra. These symmetry defects give rise to twisted sectors of codimension-two defects that are either regular or irregular corresponding to simple or higher order poles of the Higgs field. In this paper, we perform a systematic study of twisted irregular  codimension-two defects generalizing 
our earlier work in the untwisted case. 	
	In a class S setup, such twisted defects engineer 4d $\cN=2$ superconformal field theories of the Argyres-Douglas type whose flavor symmetries are (subgroups of) non-simply-laced Lie groups.
We propose formulae for the conformal and flavor central charges of these twisted theories, accompanied by nontrivial consistency checks. We also identify the 2d chiral algebra (vertex operator algebra) of a subclass of these theories and determine their Higgs branch moduli space from the associated variety of the chiral algebra.

 }

\vfill

\end{titlepage}

\eject

\tableofcontents

\section{Introduction}

The six dimensional $(2,0)$ superconformal theories (SCFT) are mysterious quantum field theories that arise either as low energy descriptions of five-branes in M-theory or in a decoupling limit of type IIB string probing ADE singularities \cite{Witten:1995zh,Strominger:1995ac,Witten:1995em}. They are rigid
strongly coupled fixed points in six dimensions that are believed to be determined by an ADE Lie algebra and have no relevant deformations that preserve the $(2,0)$ supersymmetry \cite{Henningson:2004dh,Cordova:2015vwa,Cordova:2016xhm}. Residing in a highly constrained structure, the richness of the $(2,0)$ theories lies in the collection of extended defects and their dynamics \cite{Witten:2009at}.  
In particular, a plethora of lower dimensional supersymmetric theories have been constructed by compactifications of the $(2,0)$ theory on manifolds with defect insertions. The sheer existence of the 6d parent has lead to many highly nontrivial predictions for the physics of the lower dimensional theories as well as dualities between ostensibly different field-theoretic descriptions. For many cases, these predictions are  verified by techniques that are accessible in the lower dimensions.
In this way, even though the $(2,0)$ SCFT itself does not have a simple field-theoretic construction that allows direct access to its dynamics,\footnote{The conformal bootstrap has proven to be an efficient tool to probe the fixed point physics. See \cite{Poland:2018epd} for an overview and in particular \cite{Beem:2015aoa} for the application of bootstrap methods to the 6d $(2,0)$ SCFT.} we can gain valuable insights by studying its daughter theories. 

Among all defects in $(2,0)$ SCFTs, the half-BPS codimension-two defects are one of the central focuses of investigation in recent years. 
They play a crucial role in the class S construction of 4d $\cN=2$ SCFTs by compactifying $(2,0)$ SCFT on a Riemann surface with punctures \cite{Witten:1997sc,Gaiotto:2009we,Gaiotto:2009hg}. In this setup, the codimension-two defects that extend in the 4d spacetime directions produce the punctures.\footnote{For this reason, we will use defect, puncture and singularity interchangeably when referring to the codimension-two defects.} They often give rise to global symmetries in the 4d theory and supply degrees of freedom that carry symmetry charges. 
As we review in section~\ref{sec:rev}, these codimension-two defects in the (2,0) SCFT can be described by the singularities of a Lie algebra valued one-form field, the Higgs field $\Phi$, on the Riemann surface. 
Furthermore, these defects come in two families: 
the regular (tame) defects corresponding to simple poles for $\Phi$, and the irregular (wild) defects associated to higher order singularities. 
Classification of the regular defects was given   in \cite{Gaiotto:2009we,Nanopoulos:2009uw,Chacaltana:2012zy} and the irregular defects were studied in \cite{Xie:2012hs,Wang:2015mra}. 
In particular the regular defects carry flavor symmetries that are subgroups of ADE Lie groups.

\begin{table}[htb]
	\begin{center}
		\begin{tabular}{ |c|c| c|c|c|c| }
			\hline
			$ \mf j $ ~&$A_{2N}$ &$A_{2N-1}$ & $D_{N+1}$  &$E_6$&$D_4$ \\ \hline
			Outer-automorphism $o$  &$\mZ_2$ &$\mZ_2$& $\mZ_2$  & $\mZ_2$&$\mZ_3$\\     \hline
			Invariant subalgebra  $\mf g^\vee$ &$B_N$&$C_N$& $B_{N}$  & $F_4$&$G_2$\\     \hline
			Flavor symmetry $\mf g$ &$C_N$&$B_N$& $C_{N}$  & $F_4$&$G_2$\\     \hline
		\end{tabular}
	\end{center}
	\caption{Outer-automorphisms of simple Lie algebras $\mfj$, its invariant subalgebra $\mfg^\vee$ and flavor symmetry $\mfg$ from the Langlands dual.}
	\label{table:outm}
\end{table}

\begin{figure}[htb]
	\centering
	\begin{minipage}{0.3\textwidth}
		\includegraphics[width=1\textwidth]{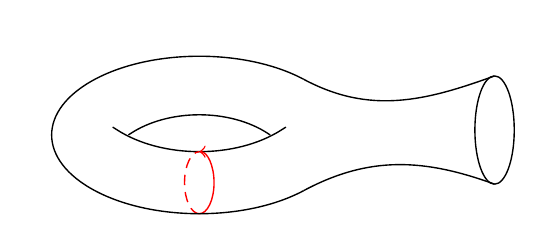}
	\end{minipage}%
	\begin{minipage}{0.1\textwidth}\begin{eqnarray*} \times~ \mR^{3,1}  \\ \end{eqnarray*}
	\end{minipage}%
		\begin{minipage}{0.3\textwidth}
		\includegraphics[width=1\textwidth]{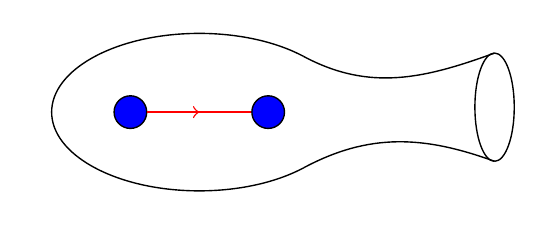}
	\end{minipage}%
		\begin{minipage}{0.1\textwidth}\begin{eqnarray*} \times~ \mR^{3,1}  \\ \end{eqnarray*}
		\end{minipage}%
	\caption{Local configurations of the codimension-one symmetry defects (red line) and codimension-two twisted  defects (blue dot) on the Riemann surface in a class S setup.}
	\label{fig:defects}
\end{figure}

The $(2,0)$ SCFT also has codimension-one defects that correspond to the discrete global symmetries associated to the outer-automorphism group of the ADE Lie algebra (see Table~\ref{table:outm}).\footnote{\label{foot:iibsym}In the IIB setup, these discrete global symmetries come from discrete isometries of the ADE singularity that preserve the hyper-K\"ahler structure  (see Table~\ref{table:sym} and \cite{Tachikawa:2011ch}). A careful reader may notice some peculiarity about the $A_{2N}$ case. For $A_{2N}$ singularity, the relevant discrete isometry is an $\mZ_4$ generated by $\sigma:(x,y,z)\to (y,-x,-z)$. But since $\sigma^2:(x,y,z)\to (-x,-y,z)$ is part of the connected $U(1)$ isometry that act by $x\to e^{i\alpha} x,y\to e^{-i\alpha} y$, combined with the fact that the $(2,0)$ SCFT has no global currents (non-R symmetry), it should act trivially on the $(2,0)$ SCFT in the decoupling limit. Therefore we only expect to see the $\mZ_2$ symmetry in the $(2,0)$ $A_{2N}$ SCFT. We thank Edward Witten for helpful discussions on this point.} Thus we can consider {\it twisted} codimension-two defects that live at the ends of these symmetry defects.\footnote{This is a higher dimensional generalization of the familiar twisted sector operators in a 2d orbifold CFT. Similarly there should also be twisted codimension-two defects in 4d $\cN=4$ super-Yang Mills theories. It would be interesting to understand their roles in S-duality and the geometric Langlands program \cite{Witten:2007td}.} On the Riemann surface in a class S setup, the codimension-one defect is represented by a twist line that either wraps a nontrivial cycle or connects a pair of twisted punctures (see Figure~\ref{fig:defects}).

 Twisted regular defects carry (subgroups of) non-simply-laced  BCFG flavor symmetry  groups and they were studied extensively in \cite{Tachikawa:2009rb,Tachikawa:2010vg,Chacaltana:2012zy,Chacaltana:2015bna,Chacaltana:2014nya,Chacaltana:2012ch}. Note that the maximal $\mZ_2$ twisted regular defects in  type $A_{2N}$ and $D_{N+1}$ $(2,0)$ SCFTs share the same flavor symmetry $USp(2N)$ but differ in Witten's global anomaly for the said symmetry \cite{Tachikawa:2011ch,Tachikawa:2018rgw}.

\begin{table}[htb]
	\begin{center}
		\begin{tabular}{ |c|c|c|c| }
			\hline
			$\mfj$~&Singularity& Symmetry & Action of generators \\ \hline
			$A_{2N}$&$x^{2N+1}+yz=0$&$\mZ_4$ & $(x,y,z)\rightarrow(-x,z,-y)$ \\ \hline
			$A_{2N-1}$&$x^{2N}+yz=0$&$\mZ_2$ & $(x,y,z)\rightarrow(-x,z,y)$ \\ \hline
			$D_{N+1}$ & $x^{N}+xy^2+z^2=0$& $\mZ_2$ & $(x,y,z)\rightarrow(x,-y,-z)$ \\ \hline
			$E_6$ & $x^4+y^3+z^2=0$& $\mZ_2$ & $(x,y,z)\rightarrow(-x,y,-z)$ \\ \hline
			$D_4$ & $x^3+y^3+z^2=0$ & $S_3$ & $(x,y,z)\rightarrow(y,x,-z)$ \\ 
			~&~&~&$(x,y,z)\rightarrow(\omega x, \omega^2 y, z)$  \\ \hline
		\end{tabular}
	\end{center}
	\caption{Discrete symmetries of ADE two-fold singularities.}
	\label{table:sym}
\end{table}

One of main purposes of this paper is to classify twisted irregular defects in $(2,0)$ SCFTs. Before we summarize the results, let us briefly recall the classification of untwisted irregular defects performed in   \cite{Xie:2012hs,Wang:2015mra} since the method we pursue here is a direct generalization.

 The classification of codimension-two defects in \cite{Xie:2012hs,Wang:2015mra} was based on analyzing consistency conditions on the higher order singularities of the Higgs field $\Phi$ on a Riemann surface with a local holomorphic coordinate $z$. After the singularity is put into the convenient semisimple form by a gauge transformation,
\ie
\Phi_z(z)= {T\over z^{2+{k\over b}}}+\dots,\quad {\rm with~} b\in \mZ^+, k\in \mZ~{\rm and}~k>-b.
\label{utphi}
\fe
where $T$ is a semisimple element of $\mfj$, the consistency condition on $\Phi$ simply says 
\ie
\Phi(e^{2\pi i} z)= \sigma_g \cdot  \Phi(z)
\label{utcon}
\fe
for some inner automorphism $\sigma_g$ of the Lie algebra $\mfj$. This puts constraints on the defining data $(T,k,b)$ of the singularity, which are then solved systematically by Kac's classification of finite order (torsion) inner automorphisms of simple Lie algebras \cite{kac2013infinite}. Generic 4d $\cN=2$ SCFTs engineered by such irregular defects are of the Argyres-Douglas (AD) type which have fractional scaling dimensions in the half-BPS Coulomb branch spectrum and are intrinsically strongly-coupled \cite{Argyres:1995jj,Argyres:1995wt,Argyres:1995xn}.
\begin{table}[!htb]
	\begin{center}
		\begin{tabular}{ |c|c|c|c| }
			\hline
			$ \mfj$& $b$  & Singularity   \\ \hline
			$A_{N-1}$&$N$ &$x_1^2+x_2^2+x_3^N+z^k=0$\\ \hline
			$~$  & $N-1$ & $x_1^2+x_2^2+x_3^N+x_3 z^k=0$\\ \hline
			
			$D_N$  &$2N-2$ & $x_1^2+x_2^{N-1}+x_2x_3^2+z^k=0$ \\     \hline
			$~$  & $N$ &$x_1^2+x_2^{N-1}+x_2x_3^2+z^k x_3=0$ \\     \hline
			
			$E_6$&12  & $x_1^2+x_2^3+x_3^4+z^k=0$   \\     \hline
			$~$ &9 & $x_1^2+x_2^3+x_3^4+z^k x_3=0$   \\     \hline
			$~$  &8 & $x_1^2+x_2^3+x_3^4+z^k x_2=0$    \\     \hline
			
			$E_7$& 18  & $x_1^2+x_2^3+x_2x_3^3+z^k=0$   \\     \hline
			$~$&14   & $x_1^2+x_2^3+x_2x_3^3+z^kx_3=0$    \\     \hline

			$E_8$ &30   & $x_1^2+x_2^3+x_3^5+z^k=0$  \\     \hline
			$~$  &24  & $x_1^2+x_2^3+x_3^5+z^k x_3=0$  \\     \hline
			$~$  & 20  & $x_1^2+x_2^3+x_3^5+z^k x_2=0$  \\     \hline
		\end{tabular}
	\end{center}
	\caption{Three-fold isolated quasi-homogeneous singularities of the cDV type corresponding to the $J^{(b)}[k]$  irregular punctures of the regular-semisimple type in \cite{Wang:2015mra}. }
	\label{table:sing}
\end{table}

A distinguished class of solutions to \eqref{utphi} and \eqref{utcon}, known as the {\it regular-semisimple} type\footnote{$T$ is a regular semisimple element of the Lie algebra $\mfj$.} gives rise to irregular codimension-two defects that are in one-to-one correspondence with three-fold quasi-homogeneous isolated singularities of the compound Du Val  (cDV) type (see Table~\ref{table:sing}).\footnote{More explicitly, the moduli space of complex structure deformations (more precisely miniversal or semiuniversal deformations) of a cDV singularity is identified with the Hitchin moduli space of the Higgs bundle on a sphere with one irregular puncture of the regular-semisimple type. This identification can also be interpreted as a correspondence between non-compact ``CY3'' integrable systems and Hitchin integrable systems in the sense of \cite{Diaconescu:2006ry} (see also \cite{beck} for a review and some recent developments). It would be interesting to make this more precise. 
 }
 This connection between the two very different types of singularities is established by the observation that identical 4d $\cN=2$ SCFTs are engineered by i) compactifying $(2,0)$ SCFT on $\mP^1$ with such a irregular defect inserted; ii) the decoupling limit of IIB string probing a cDV singularity. We review the general untwisted irregular defects, the resulting 4d $\cN=2$ SCFTs as well as their physical data in section~\ref{sec:rev}.

In this paper, we incorporate outer-automorphism twists into the configurations of codimension-two irregular defects in the $(2,0)$ SCFT.
The consistency condition for the Higgs field singularity at $z=0$ on the Riemann surface \eqref{utcon} is modified to 
\ie
\Phi(e^{2\pi i} z)=\sigma_g o \cdot \Phi(z)
\fe
where $\sigma_g o$ labels an outer-automorphism of $\mfj$ with $o$ generating automorphisms of the Dynkin diagrams (see Table~\ref{table:outm}) and the parameter $b$ in \eqref{utphi} is replaced by $b_t$. As we explain in section~\ref{sec:tip}, this constraint can be solved by invoking the classification of  finite order (torsion) outer-automorphisms of simple Lie algebras, which is also given  in \cite{kac2013infinite}. Restricting the polar matrix $T$ in \eqref{utphi} to be regular semisimple again gives rise to a distinguished class of twisted irregular defects, which can be put into the three-fold form in Table~\ref{table:SW}.\footnote{We emphasize here that, although suggestive, we do not yet understand the physical meaning of the singular geometries of Table~\ref{table:SW} in type IIB string theory as they involve branch cuts. Nevertheless, for practical purposes, we find them to be a useful mnemonic for the 4d SCFTs engineered by such twisted irregular defects (along with another twisted regular defect).
}
We also explicitly identify the continuous free parameters of these defects with flavor symmetry masses and exactly marginal couplings.
\begin{table}[!htb]
\begin{center}
\begin{tabular}{|c|c|c|c|}
\hline
$\mfj$ with twist & $b_t$ & SW geometry at SCFT point & $\Delta[z]$ \\[1pt] \hline
 $A_{2N}/\mZ_2$ & $4N+2$ &$x_1^2+x_2^2+x^{2N+1}+z^{k'+{1\over2}}=0$ & ${4N+2\over 4N+2k'+3}$  \\[3pt] \hline
 ~& $2N$ & $x_1^2+x_2^2+x^{2N+1}+xz^{k'}=0$ & ${2N\over k'+2N}$ \\[3pt] \hline
  $A_{2N-1}/\mZ_2$& $4N-2$ & $x_1^2+x_2^2+x^{2N}+xz^{k'+{1\over2}}=0$ & ${4N-2\over 4N+2k'-1}$ \\[3pt] \hline
~ & $2N$ &$x_1^2+x_2^2+x^{2N}+z^{k}=0$ & ${2N\over 2N+k'}$  \\[3pt] \hline
 $D_{N+1}/\mZ_2$& $2N+2$ & $x_1^2+x_2^{N}+x_2x_3^2+x_3z^{k'+{1\over2}}=0$ & ${2N+2\over 2k' +2N+3}$ \\[3pt] \hline
~ & $2N$ &$x_1^2+x_2^{N}+x_2x_3^2+z^{k'}=0$ & ${2N\over k'+2N}$  \\[3pt] \hline
 $D_4/\mZ_3$ & $12$ &$x_1^2+x_2^{3}+x_2x_3^2+x_3z^{k'\pm {1\over3}}=0$ & ${12\over 12+3k'\pm1}$  \\[3pt] \hline
  ~& $6$ &$x_1^2+x_2^{3}+x_2x_3^2+z^{k'}=0$ & ${6\over 6+k'}$  \\[3pt] \hline
$E_6/\mZ_2$& $18$ &$x_1^2+x_2^{3}+x_3^4+x_3z^{k'+{1\over2}}=0$ & ${18\over 18+2k'+1}$  \\[3pt] \hline
  ~& $12$ &$x_1^2+x_2^{3}+x_3^4+z^{k'}=0$ & ${12\over 12+k'}$  \\ [3pt]\hline
~ & $8$ &$x_1^2+x_2^{3}+x_3^4+x_2z^{k'}=0$ & ${8\over 12+k'}$  \\[3pt] \hline
\end{tabular}
\caption{Seiberg-Witten geometry of twisted theories at the SCFT point.}
\label{table:SW}
\end{center}
\end{table}

Moving onto section~\ref{sec:tt}, we classify 4d $\cN=2$ SCFTs that are engineered by such twisted irregular defects in class S setups which we refer to as  {\it twisted theories}. The twisted theories come in infinite families (labelled by $b_t$) for each choice of the simple Lie algebra $\mfj$ (see Table~\ref{table:SW}). We spell out a simple procedure to extract physical data of such theories from our descriptions and propose formulae for the conformal and flavor central charges. Most of the theories we construct here are new but we  identify in our setups several (sequences of) known constructions that only involve regular defects. They provide a nontrivial consistency check of our construction and central charge formulae. 
 
A general 4d $\cN=2$ SCFT  is known to have a nontrivial protected sector described by a 2d chiral algebra \cite{Beem:2013sza}. Recent developments indicate that it captures information of both Coulomb branch and Higgs branch physics of the 4d theory \cite{Cordova:2015nma,Buican:2015ina,Cecotti:2015lab,Xie:2016evu,Creutzig:2017qyf,Fredrickson:2017yka,Song:2017oew,Beem:2017ooy,Kozcaz:2018usv}. In particular, the Higgs branch of the 4d theory is identified with the associated variety of the 2d chiral algebra \cite{arakawa2015associated}. In section~\ref{sec:voa}, we identify the associated 2d chiral algebra or vertex operator algebra (VOA) for a subclass of the twisted theories and determine the Higgs branch of these theories from 
the associated variety of the VOA.  
We end in section~\ref{sec:sum}. with a summary and future directions.

%%%%%%%%%%%%%%%%%%%

\section{Background review}
\label{sec:rev}
In this section, we review the description of codimension-two BPS defects in 6d $(2,0)$ SCFTs in terms of the Higgs field  and explain the class S construction of 4d $\cN=2$ SCFTs using the  Hitchin system on a Riemann surface with defect insertions. We also summarize the classification of untwisted defects in the last subsection. Readers familiar with these topics may safely skip this section.

\subsection{Hitchin system and 4d $\cN=2$ SCFTs}
A large class of 4d $\cN=2$ superconformal field theories can be engineered by the twist compactification of the 6d $(2,0)$ type $J=ADE$ theories on a Riemann surface $\cC$, usually referred to as the UV curve or Gaiotto curve \cite{Witten:1997sc,Gaiotto:2009we}. The Riemann surface $\cC$ can come with punctures at isolated points $\{p_i\}$, which correspond to codimensional-two BPS defects in the 6d SCFT. In M-theory, the $A_N$ type (2,0) SCFT captures the low energy dynamics of a stack of fivebranes and the codimensional-two defect is described by another stack of fivebranes which share four longitudinal directions with the former. To produce $4d$ theories with $\cN=2$ supersymmetry, we have the first stack of fivebranes wrapping $\cC$, whereas the defect fivebranes extend along the cotangent fibers of $\cC$ at the singularities $p_i$.
\ie	\hspace*{-4cm}  
\begin{blockarray}{ccccccc}
	{     \mbox{\textrm{Defect fivebranes:}}} &  \mR^{2,1}~\times &S^1   &\times& T^*_{p_i}\cC & &&& \\
	&  &  & & \cap \\
	{\mbox{\rm M-theory:}}&  \mR^{2,1}  ~\times &S^1  & \times & T^* \cC & \times   \rlap{$\mR^3$} &\\
	& &  & & \cup \\
	{\mbox{\rm 6d (2,0) SCFT on fivebranes:}}& \mR^{2,1}  ~\times &S^1 & \times &  \cC & \\    
	%  \llap{\hspace{-10\arraycolsep} twist~compactification~~$\swarrow$~~$\searrow$ twist~compactification}  
	&\llap{\phantom{0}\phantom{0}\hspace{ 3\arraycolsep} $\swarrow$  }  
	&& \rlap{\phantom{0}\phantom{0}\hspace{ 3\arraycolsep}  $\searrow$ }  
	\\
	& \llap{\hspace{ 30\arraycolsep} 4d  $\cN=2$ SCFT $\cT_4[\cC]$:~$ \mR^{2,1}~\times~S^1$}
	&     
	\rlap{\hspace{20\arraycolsep} 5d  $\cN=2$ SYM:~$ \mR^{2,1}~\times~\cC$}
	&   &  & & 
	\\
	&\llap{\phantom{0}\phantom{0}\hspace{ 3\arraycolsep} $\searrow$  }  
	&& \rlap{\phantom{0}\phantom{0}\hspace{ 3\arraycolsep}  $\swarrow$ }  
	\\
	&\rlap{\hspace{-20\arraycolsep} 3d  $\cN=4$ SCFT $\cT_3[ \cC]$:~$ \mR^{2,1}$} & &&\\
\end{blockarray}
\label{m5setup}
\fe
To decode information about the 4d $\cN=2$ SCFT from this construction where the 6d parent has no explicit description (e.g. in terms of a Lagrangian), it is useful to consider the alternate compactification of the 6d theory on a circle transverse to the Riemann surface $\cC$. The resulting 5d theory is believed to be the $\cN=2$ super Yang-Mills (MSYM) with  gauge group $J$  (up to higher derivative corrections) \cite{Seiberg:1996bd, Seiberg:1997ax,Douglas:2010iu,Lambert:2010iw }.

Upon twisted compactification of the 5d theory on the Riemann surface $\cal C$ with holomorphic coordinate $z$, there's a natural principal $J$-bundle $E$ over $\cal C$ with gauge connection $A=A_z dz+A_{\bar z}d\bar z$ and two of the five 5d scalars  combine into a $(1,0)$-form  $\Phi=\Phi_z dz$ valued in the adjoint bundle ${\rm ad}(E)$. The supersymmetric configurations of the twisted theory are governed by the Hitchin equations
\ie
&F+ [\Phi,\bar\Phi]=0,
\\
&\bar\partial_A \Phi=d\bar z(\partial_{\bar z}\Phi+[A_{\bar z},\Phi])=0,
%\\
%\partial_A \bar\Phi=d z(\partial_{ z}\bar\Phi+[A_{  z},\bar\Phi])&=0,
\label{hitchin}
\fe 
where $F$ denotes the curvature two-form of $A$. The pair $(E,\Phi)$ subject to \eqref{hitchin} is referred to as the Higgs bundle and $\Phi$ is the Higgs field. In particular, the second line of \eqref{hitchin} implies $\Phi$ is a holomorphic section of  $\cK \otimes {\rm ad}(E)$ where $\cK$ denotes the canonical bundle on $\cC$. For a fixed structure group $J$, the moduli space of Higgs bundles over $\cC$ (solutions to Hitchin equations \eqref{hitchin} modulo gauge redundancy) corresponds to the Hitchin moduli space $\cM_{\rm H}(\cC)$.

The twisted compactification of 5d MSYM leads to a particular description of the 3d $\cN=4$ SCFT in the IR, with a Higgs branch identical to $\cM_{\rm H}(\cC)$ thanks to supersymmetry. Meanwhile the other order of compactifications in \eqref{m5setup} (first on $\cC$ then $S^1$) provides another description of the same 3d SCFT related by mirror symmetry, where $\cM_{\rm H}(\cC)$ now describes the Coulomb branch. 
\ie
\begin{array}{rcl}
	\text{Coulomb branch of $\cT_4[\cC]$ on $S^1$}&\xleftrightarrow{~~\text{\small 3d mirror symmetry}~~} &\text{Higgs branch of 5d MSYM on $\cC$}\\
	%\\
	%$\rotatebox[origin=c]{-45}{$\longleftrightarrow$}$   &  \neswarrow & $\rotatebox[origin=c]{45}{$\longleftrightarrow$} $ \\
	%\\
	\nwsearrow      & &   \neswarrow    \\
	\rlap{\hspace{-7\arraycolsep} \text{Hitchin moduli space $\cM_{\rm H}(\cC)$}}&
\end{array}.
\label{}
\fe 

$\cM_{\rm H}(\cC)$ is  a hyper-K\"ahler manifold of rich structure that encodes dynamics of the 4d theory $\cT_4[\cC]$ as well as its 3d descendant. In one complex structure, $\cM_{\rm H}(\cC)$ is equivalent to the moduli space of $\mf j_{\mC}$-valued flat connections $\cA=A+i\Phi+i\bar{\Phi}$ on $\cC$. For example when $\cC$ has genus $g$ with no punctures,  $\cM_{\rm H}(\cC)$ is parametrized by the holonomies around the $2g$ cycles labelled by elements $U_{1\leq i\leq g}$ and $V_{1\leq i\leq g}$ in  $J_{\mC}$  that are  subject to the constraint
\ie
1=U_1V_1 U_1^{-1} V_1^{-1}\dots U_gV_g U_g^{-1} V_g^{-1},
\label{cs1}
\fe
and modulo $J_{\mC}$ gauge transformations. In particular, $\cM_{\rm H}(\cC)$ has complex dimension
\ie
\dim \cM_{\rm H}(\cC)=(2g-2)\dim J.
\fe
In another complex structure,\footnote{The two complex structures relevant for \eqref{cs1} and \eqref{cs2} are usually referred to as the $J$ and $I$ complex structures respectively in the literature \cite{Kapustin:2006pk}.} $\cM_{\rm H}(\cC)$ exhibits a natural fibration structure
\ie
\pi:\cM_{\rm H}(\cC)\rightarrow {\cB}
\label{cs2}
\fe
by taking Casimirs of $\Phi$ (gauge invariant differentials), known as the Hitchin fibration \cite{Hitchin:1986vp,Hitchin:1987mz} 
The base manifold (a complex affine space) $\cB$ is special K\"ahler and identified with the Coulomb branch of the 4d theory $\cT_4[\cC]$, while the fiber is (generically) a complex torus of dimension ${1\over 2}\dim \cM_{\rm H}(\cC)$ and corresponds to the electric and magnetic holonomies\footnote{In the 3d perspective, they are parametrized by the additional (dual) scalars in the dimensional reduced vector multiplet and are part of the moduli space.} on $S^1$.
In this complex structure, there is a holomorphic symplectic form $\Omega_I$  which defines a Poisson bracket for functions on $\cM_{\rm H}(\cC)$. Hence the Hitchin moduli space $\cM_{\rm H}(\cC)$ becomes a complex integrable system with commuting Hamiltonians given by the Casimirs of $\Phi$ and the fibers of $\cM_{\rm H}(\cC)$ are identified with the orbits of the Hamiltonian flows which are special Lagrangian with respect to $\Omega_I$ \cite{Donagi:1995cf}.

We can recover the usual Seiberg-Witten (SW) description \cite{Seiberg:1994rs,Seiberg:1994aj} of the low energy dynamics of the 4d theory $\cT_4[\cC]$ as follows. The SW curve $\Sigma$ governing the Coulomb branch dynamics is equivalent to the spectral curve of the Hitchin system,\footnote{The sheets of the SW curve $\Sigma$ viewed as a branched cover of $\cC$ are labelled by eigenvalues of $\Phi_z$.}
\ie
\det(x dz-\Phi)=0,
\fe
and the SW differential is identified with $\lambda=xdz$. More explicitly, the SW curves can be put in the following forms
\ie
& A_{N-1}:~ x^N+\sum_{i=2}^N \phi_i(z) x^{N-i}=0,  
\\
& D_{N}:~x^{2N}+\sum_{i=1}^{N-1} {\phi}_{2i}(z)x^{2N-2i}+(\tilde{\phi}_N)^2=0, 
\\
& E_6:~\phi_2(z), \phi_5(z), \phi_6(z), \phi_8(z), \phi_9(z), \phi_{12}(z),       
\\
& E_7:~  \phi_2(z), \phi_6(z), \phi_8(z), \phi_{10}(z), \phi_{12}(z), \phi_{14}(z), \phi_{18}(z),       
\\
&  E_8:~  \phi_2(z), \phi_8(z), \phi_{12}(z), \phi_{14}(z), \phi_{18}(z), \phi_{20}(z),   \phi_{24}(z),  \phi_{30}(z).            
\fe
Here $\phi_i(z)$ is a degree $i$ differential on $\cC$ and they generate the ring of fundamental invariants (Casimirs) of the Lie algebra. For $E_N$ case, we only list the independent differentials. The coefficients $u_{i,j}$ of these differentials in the $z$ expansion
\ie
\phi_i(z)=\sum_{j} u_{i,j} z^j,
\label{epsw}
\fe
encode the 
Coulomb branch parameters of the theory.\footnote{Some of these coefficients are redundant (unphysical) and can be fixed by a coordinate change of $(x,z)$ that preserves the SW differential $\lambda$ up to an exact term.} 

%
%
%Expanding the SW curve as
%\ie
% \sum_{(i,j)\in S} u_{i,j} x^i z^j=0,
%\label{epsw}
%\fe
%%\ie
%%\sum_{i=1}^r x^{m_i} \epsilon_{m_i}(z)=0 
%%  \label{epsw}
%%\fe
%%where $\{m_i\}$ are the degrees of fundamental invariants of $J$ (see Table~\ref{}). 
%the coefficients $u_{i,j}$ correspond to parameters of the 4d Coulomb branch.
Since the integral of the SW differential along a one-cycle gives the mass of a BPS particle, the SW differential $\lambda=x dz$ has scaling dimension one and consequently 
\ie
&\Delta [x]+\Delta [z]=1.
\fe
This allows us to determine the scaling dimensions of $u_{i,j}$ by demanding each term of \eqref{epsw} to have the same scaling dimensions\footnote{This is meaningful since the theory $\cT_4[\cC]$ is assumed to be conformal.}: relevant chiral couplings are given by those with $\Delta [u_{i,j}]<1$, Coulomb branch operators   if $\Delta [u_{i,j}] > 1$ and masses if $ \Delta [u_{i,j}]=1$.\footnote{There are two caveats to this prescription: 
	(i) $u_{i,j}$ with integral 
	scaling dimension  $\Delta[u_{i,j}]\geq 2$ may correspond to a homogeneous polynomial (flavor Casimir) in the mass parameters; (ii) there can be constraints among $u_{i,j}$. Both subtleties can be taken care of systematically from information of the puncture(s) on $\cC$ \cite{Chacaltana:2012zy}.}

\subsection{Review of untwisted (ir)regular defects}

The relevant codimension-two defects in the 6d $(2,0)$ theory of type $\mf j$ can be characterized by singular boundary conditions for the Higgs field $\Phi$ on $\cC$. Supposing the defect is located at $z=0$ on $\cC$, it is convenient to perform a gauge transformation (that may involve fractional powers of $z$) to put $\Phi_z$ in the following form:
\ie
\Phi_z= \sum_{k\geq \ell\geq -b} {T_\ell\over z^{2+\ell/b}}+\dots,\quad 
b\in \mZ^+, k\in \mZ
%,~k\geq -b
\label{uts}
\fe
where each $T_\ell$ is a semisimple element  of ${\mf j}$. In other words, by a gauge transformation, we have $T_\ell\in \mfh$, a Cartan subalgebra of $\mfj$, for all $\ell$ \cite{Witten:2007td}.
We have suppressed the non-divergent terms in \eqref{uts}.\footnote{We emphasize here that from the perspective of the resulting 4d theory, the coefficient matrices $\{T_\ell \}$ of the polar terms correspond to parameters such as chiral couplings and masses whereas the Coulomb branch moduli are encoded in the non-divergent terms of \eqref{uts}.}   To ensure the Higgs field $\Phi$ is well-defined on $\cC$, we require
\ie
\Phi(e^{2\pi i} z)=
%\sigma_g \cdot\Phi(z)\equiv 
g \Phi(z) g^{-1}~\Rightarrow~ g  T_\ell  g^{-1}= \omega^\ell T_\ell 
\label{utgi}
\fe
for some $g\in J$ and $\omega=e^{2\pi i\over b}$.

The usual regular (tame) punctures correspond to $b=-k$ in which case $\Phi$ has a simple pole (go to a branch cover if necessary)
\ie
\Phi_z=  {T \over z }+ \dots
\label{ruts}
\fe
and the constraint \eqref{utgi} is trivialized by taking $g=1$. The regular punctures are thus classified by the conjugacy classes of $T$. When $\mfj$ is a  classical Lie algebra, these  conjugacy classes are labelled by the Hitchin partitions which are related to the Nahm partitions by the Spaltenstein map. The generalization to exceptional Lie algebras involves the Bala-Carter labels \cite{Chacaltana:2012zy}.  The flavor symmetry associated to the puncture can be directly read off from the Nahm label and the entries of $T$ correspond to the mass parameters. The local contributions to the conformal and flavor central charges can also be computed systematically \cite{Chacaltana:2012zy}.

The irregular (wild) punctures arise when $k>-b$ where $\Phi$ has a higher order pole.\footnote{The Hitchin pole in the original holomorphic form (as opposed to  \eqref{uts}) will have higher integral pole orders but the coefficient matrices are not necessarily semisimple \cite{Witten:2007td}.} Now \eqref{utgi} puts nontrivial constraints on $T_\ell$ and $b$ which were solved by Kac \cite{kac_1990}.\footnote{Here we assume that the Hitchin pole is irreducible. In other words, the structure group $J$ associated to the Higgs bundle is not reduced.} One can associate to $g$ an inner automorphism of order $b$ (torsion automorphism) $\sigma_g$ of $\mf j$ which then introduces a grading on $\mf j$
\ie
{\mf j}=\bigoplus_{m \in \mZ/b\mZ} \mf j^m.
\fe
In particular $T_\ell$ is a semisimple element in $\mf j^\ell$. Such finite-order inner automorphisms are classified by Kac (see \S 8 of the textbook \cite{kac_1990}).\footnote{See \cite{Xie:2017vaf,Xie:2017aqx} for a recent discussion relevant for 4d $\cN=2$ SCFTs.}  

\begin{figure}[htb]
\centering
	\begin{minipage}{0.3\textwidth}
		\includegraphics[width=.8\textwidth]{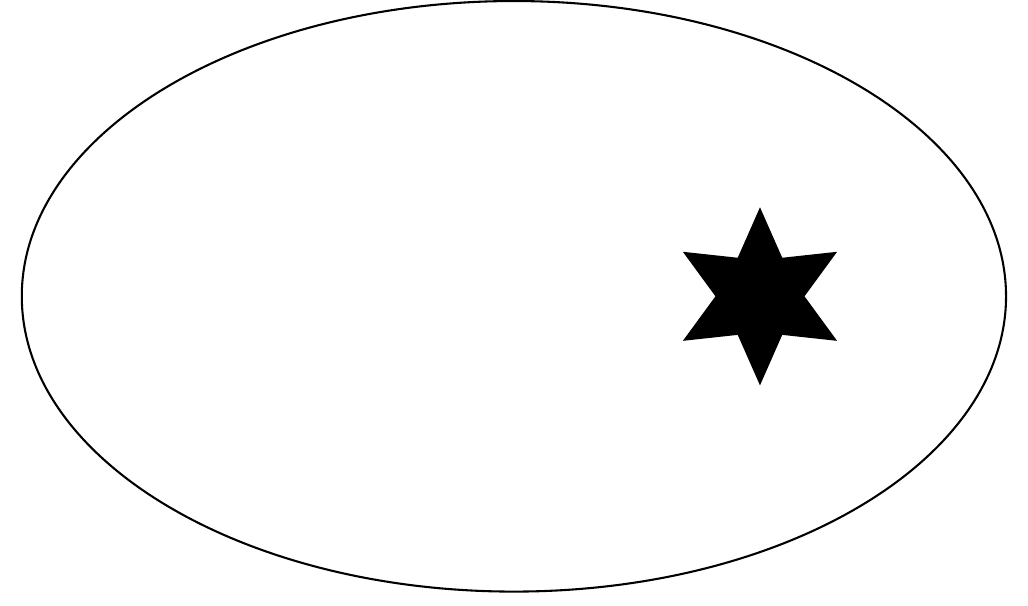}
	\end{minipage}%
\begin{minipage}{0.1\textwidth}
\quad
		\end{minipage}%
		\begin{minipage}{0.3\textwidth}
		\includegraphics[width=.8\textwidth]{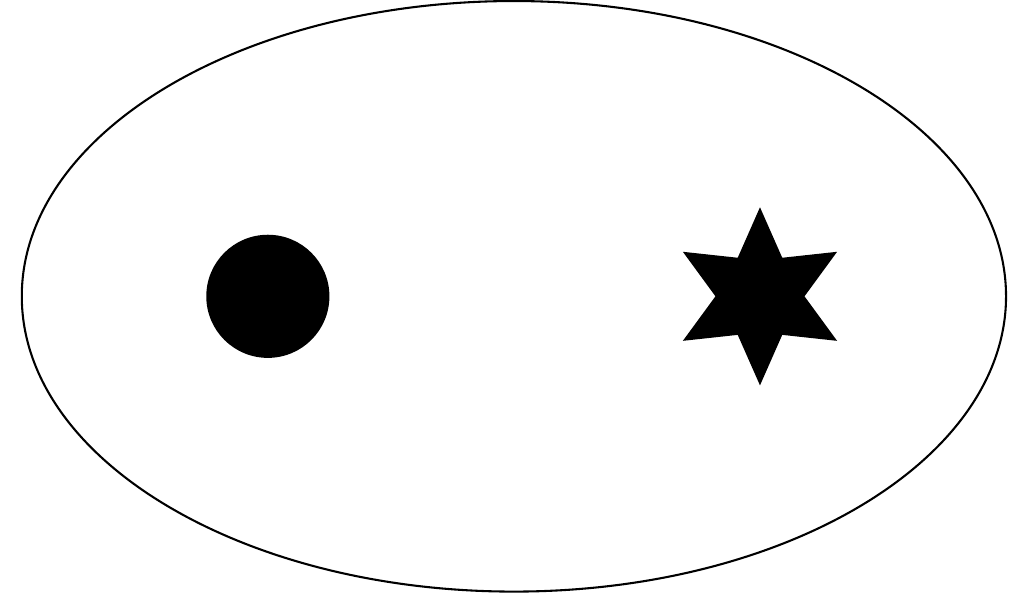}
	\end{minipage}%
%		\begin{minipage}{0.3\textwidth}
%			\includegraphics[width=.8\textwidth]{tsp.pdf}
%		\end{minipage}%
\caption{The  class S setup that involves one irregular defect (star) with or without one regular defect (dot) on a sphere. They engineer $J^{(b)}[k]$ and $(J^{(b)}[k],Y)$ theories in \cite{Wang:2015mra} respectively. Here $J^{(b)}[k]$ labels the irregular defect and $Y$ labels the regular defect.}
			\label{fig:utsp}
\end{figure}

Since we are interested in 4d $\cN=2$ SCFTs, the configurations of defects on the Riemann surface $\cC$ must be consistent with a $U(1)_R$ symmetry.  
In the absence of irregular punctures on $\cC$, the $U(1)_R$ generator is identified with the rotation generator $R_{45}$ for the $SO(5)$ R-symmetry group of the 5d  MSYM that acts on the Higgs field as $[R_{45},\Phi]=\Phi$. The regular codimension-two defects are conformal\footnote{The more precise statement is that the boundary condition that defines a regular defect \eqref{ruts} is scale-invariant. In 4d we expect this further implies conformal invariance.} thus the number and positions of the regular punctures on $\cC$ as well as the topology of $\cC$ are unconstrained.\footnote{This is true as long as the Hitchin moduli space $\cM_{\rm H}(\cC)$ has a non-negative dimension. For example, one cannot have three simple punctures on a $\mP^1$ in the $A_{N>1}$ $(2,0)$ SCFT.} On the other hand, in the presence of an irregular defect defined by \eqref{uts}, the potential 4d $U(1)_R$ generator involves a combination of $R_{45}$ and $U(1)_z$ that acts as
\ie
U(1)_R \subset SO(2)_{45} \times U(1)_z:~ \Phi\to e^{i\A}\Phi,\quad z \to e^{i \A {b\over k+b}} z.
\fe
so that the leading polar matrix $T_k$ in \eqref{uts} is preserved. 
For this to be globally defined on $\cC$, we are restricted to consider $\cC=\mP^1$ with either a single irregular puncture, or an irregular puncture accompanied by a regular puncture, located at the two fixed points $z=0,\infty$ of $U(1)_z$ (see Figure~\ref{fig:utsp}).

A distinguished class of irregular punctures \eqref{uts} of the {\it regular semisimple} type in general 6d (2,0) SCFTs were classified in \cite{Wang:2015mra}, generalizing earlier work for $A$-type (2,0) theories \cite{Xie:2012hs}. The Hitchin pole in these cases is characterized by a {regular} semi-simple element $T_k$ in $\mf j$. The commutant of $T_k$ in $\mfj$  is a Cartan subalgebra $\mfh \ni T_k$. Restricted to $\mfh$,  the inner automorphism $\sigma_g$ with a regular eigenvector $T_k$ corresponds to a   regular element  in the Weyl group $W(\mfh)$ 
in Springer's classification of regular elements for complex reflection groups \cite{springer1974regular}.   
The orders of these  regular elements (known as regular numbers $d$ in \cite{springer1974regular}) are listed in Table~\ref{table:bv1} and each order is associated to an unique regular element up to conjugation by $W(\mfh)$.  

\begin{table}[!htb]
	\begin{center}
		\begin{tabular}{ |c|c| }
			\hline
			$\mfj $ & $d$ (is divisor of)  \\ \hline
			$A_{N-1}$& $N,~N-1$ \\ \hline
%			$B_N$   & $2N$  \\     \hline
%			$C_N$   & $2N$  \\     \hline
			$D_N$   & $2N-2,~N$  \\     \hline
			$E_6$   & $12,~9,~8 $ \\     \hline
			$E_7$   & $18,~14$  \\     \hline
			$E_8$   & 30,~24,~20 \\     \hline
%			$G_2$   & $6 $ \\     \hline
%			$F_4$   & $12,~8$  \\     \hline
					\end{tabular}
	\end{center}
	\caption{Regular numbers (orders of regular semisimple elements) of Weyl groups.}
	\label{table:bv1}
\end{table}

The corresponding torsion automorphisms have the same orders except for $\mfj=A_{2N}$ and $d=2N-1$ in which case $\sigma_g$ has order $4N-2$.\footnote{\label{footnote:A2norder}In this case, since the Weyl group $W(\mfh)$ acts faithfully on $T_\ell\in \mfh$ we can effectively identify $b$ with $d$ as opposed to the order of $\sigma_g$ which is twice as big.}  
 Each regular element induces a grading of its commutant Cartan subalgebra
 \ie
 {\mf h}=\bigoplus_{m \in \mZ/d\mZ} \mf h^m.
 \fe
 where $\mfh^1$ contains the regular semisimple element. 
The corresponding irregular Hitchin pole is determined by \footnote{Recall that $T_k=T_1$ here since the subscript takes value mod $b$. \label{footnote:idip}}
\ie
\boxed{
b=d,~\gcd(k,b)=1,~{\rm and~} T_m\in \mfh^m~{\rm with~} T_1 {\rm~regular~semisimple}.
}
\fe

This class of irregular punctures and the resulting 4d theories were labelled by $J^{(b)}[k]$ in \cite{Wang:2015mra} where $b$ takes the values of the regular numbers  in Table~\ref{table:bv1} including their divisors.\footnote{The associated torsion inner automorphism has order $b\over \gcd(k,b)$ (see around Footnote~\ref{footnote:A2norder} for a caveat for $A_{2N}$). We could have restricted to $\gcd(k,b)=1$ in the regular semisimple case and consider all possible values of $b$ (including the divisors of the entries in Table~\ref{table:bv1}) when defining the $J^{(b)}[k]$ punctures. However in writing down formulae (e.g. for central charges), we find it more convenient to allow $\gcd(k,b)\neq 1$ while keeping in mind the identification $J^{(b)}[k]=J^{(b')}[k']$ for $b'={b\over \gcd(b,k)},~k'={k\over \gcd(b,k)}$. }  
The Coulomb branch spectrum and conformal central charges were computed in \cite{Wang:2015mra}, where it was shown that these theories are in one to one correspondence with those constructed by IIB string probing three-fold isolated quasi-homogeneous singularities of compound Du Val (cDV) type (see Table~\ref{table:sing}). The case of including a regular puncture with nontrivial flavor symmetry was also considered in \cite{Wang:2015mra}. In particular a class of theories denoted by  $(J^{(b)}[k],F)$ with nonabelian ADE flavor symmetry were constructed by including an additional maximal (full) regular puncture. A nice feature of these  $J^{(b)}[k]$ and $(J^{(b)}[k],F)$  theories is that many of them have simple 2d chiral algebras that correspond to either W-algebra minimal model or affine Kac-Moody algebra.
\begin{table}[!htb]
	\begin{center}
		\begin{tabular}{ |c|c|c|c| }
			\hline
			$ \mfj$& Singularity & $b$   \\ \hline
			$A_{N-1}$ &$x_1^2+x_2^2+x_3^N+z^k=0$&$N$\\ \hline
			$~$& $x_1^2+x_2^2+x_3^N+x_3 z^k=0$ & $N-1$\\ \hline
			
			$D_N$   & $x_1^2+x_2^{N-1}+x_2x_3^2+z^k=0$&$2N-2$ \\     \hline
			$~$   &$x_1^2+x_2^{N-1}+x_2x_3^2+z^k x_3=0$ & $N$\\     \hline
			
			$E_6$  & $x_1^2+x_2^3+x_3^4+z^k=0$&12   \\     \hline
			$~$  & $x_1^2+x_2^3+x_3^4+z^k x_3=0$ &9  \\     \hline
			$~$  & $x_1^2+x_2^3+x_3^4+z^k x_2=0$  &8   \\     \hline
			
			$E_7$  & $x_1^2+x_2^3+x_2x_3^3+z^k=0$& 18   \\     \hline
			$~$  & $x_1^2+x_2^3+x_2x_3^3+z^kx_3=0$ &14    \\     \hline

			$E_8$   & $x_1^2+x_2^3+x_3^5+z^k=0$&30   \\     \hline
			$~$   & $x_1^2+x_2^3+x_3^5+z^k x_3=0$ &24   \\     \hline
			$~$   & $x_1^2+x_2^3+x_3^5+z^k x_2=0$ & 20  \\     \hline
		\end{tabular}
	\end{center}
	\caption{Three-fold isolated quasi-homogeneous singularities of cDV type corresponding to the $J^{(b)}[k]$  irregular punctures of the regular semisimple type in \cite{Wang:2015mra} . }
	\label{table:sing}
\end{table}
In particular the flavor central charge of the $(J^{(b)}[k],F)$ theory \cite{Xie:2013jc,Xie:2016evu,Xie:2017vaf,Xie:2017aqx}:
\ie
k_F=h-{b\over b+k}
\fe
was identified with the level of the affine Kac-Moody algebra $\widehat \mfj$ by
$k_{2d}=-k_F$. In section~\ref{sec:voa}, we will see how the twisted theories realize the other types of affine Kac-Moody algebra and W-algebra associated to non-simply-laced simple Lie algebras.

%%%%%%%%%%%%%%%%%%%%%%
 
\section{Twisted irregular defects}
\label{sec:tip}

In this section, we study the classification of general twisted codimension-two BPS defects in 6d $(2,0)$ SCFTs. Along the way, we make connection to known results in the literature. We then describe in detail the classification of a distinguished class of twisted irregular defects of the {\it regular-semisimple} type. The physical interpretation for the parameters of the twisted irregular defect is explained in the last subsection.

\subsection{Classification of general twisted defects}
When $\mf j$ has a nontrivial outer-automorphism group $\rm{Out}(\mf j)$, we can decorate the puncture (defect) with a monodromy twist by  $o \in {\rm Out}(\mf j)$ around the singularity (see Table~\ref{table:outm}). 
In the IIB realization of the $(2,0)$ SCFT from the decoupling limit of string probing an ADE singularity, the above outer-automorphism group arises from the symmetry of the singularity \cite{slodowy2001simple} (see Footnote~\ref{foot:iibsym} and Table~\ref{table:sym}).

Locally the Higgs field behaves as
\ie
\Phi_z= \sum_{k\geq \ell} {T_\ell\over z^{2+\ell/b_t}}+\dots,\quad 
b_t\in \mZ^+, k\geq -b_t\in \mZ
%,~k\geq -b
\label{ts}
\fe
and the twist amounts to modifying the requirement \eqref{utgi}  to
%\ie
%\Phi(e^{2\pi i} z)=g[o(\Phi(z))]g^{-1}
%\fe
\ie
\Phi(e^{2\pi i} z)=
%\sigma_go\cdot \Phi(z)\equiv
 g[o(\Phi(z))]g^{-1}~\Rightarrow~ g  [o(T_\ell)]  g^{-1}= \omega^\ell T_\ell 
\label{tgi}
\fe
for some $g\in J/{  G}^\vee$ where ${G}^\vee$ is the invariant subgroup of $J$ with respect to $o$ \cite{Chacaltana:2012zy}. 
Globally the twisted punctures must come in pairs connected by twist lines.  

Compared to the untwisted case, the gauge transformation required for the Higgs field to be well-defined is now $\sigma_g o $ which is a {\it twisted  torsion automorphism} of $\mfj$ of order $b_t$ that projects to a nontrivial element in ${\rm Out}(\mfj)$.\footnote{In particular $b_t$ is always a multiple of $|o|$.} It induces a grading on the Lie algebra
\ie
{\mf j}=\bigoplus_{m \in \mZ/b_t\mZ} \mf j^m.
\label{tg}
\fe
such that $T_\ell$ is a semisimple element of $\mfj^\ell$ which has eigenvalue $\omega^\ell$ under $\sigma_g o$.\footnote{In the superscript of $\mfj^\ell$, $\ell$ is understood to be mod $b_t$.} 
The twisted torsion automorphisms for simple Lie algebras were classified in \cite{kac_1990}. 

A subclass of twisted punctures with leading simple pole  are called regular twisted punctures in which case the constraint \eqref{tgi} is solved by $\sigma_g=1$ and
\ie
\Phi_z={T_{-2}\over z}+{T_{-1} \over z^{1/2}}+T_0+\dots, \quad T_{-2},T_{0}\in \mfj^0~{\rm and}~ T_{-1} \in \mfj^1 
\fe
for $o$ of order 2 in the case of $\mfj=A_n,D_n,E_6$
and
\ie
\Phi_z={T_{-3}\over z}+{T_{-2} \over z^{2/3}}+{T_{-1} \over z^{1/3}}+T_0+\dots, \quad T_{-3},T_{0}\in \mfj^0,~ T_{-2} \in \mfj^2 
~{\rm and}~ T_{-1} \in \mfj^1 
\fe
for $o$ of order 3 in the case of $\mfj=D_4$. 
Various physical information associated to these twisted punctures were studied in \cite{Chacaltana:2012zy} including the local contributions to the Coulomb branch and Higgs branch, flavor symmetry and central charges. The 4d class S theories constructed from these punctures were studied extensively \cite{Chacaltana:2012ch,Chacaltana:2013oka,Chacaltana:2014nya,Chacaltana:2015bna,Chacaltana:2016shw}.

More generally, we have an twisted irregular puncture from \eqref{ts}. In the presence of outer-automorphism twist, to construct 4d $\cN=2$ SCFT with one twisted irregular  puncture, we must pair it with a regular twisted puncture on $\mP^1$.

\subsection{Twisted irregular defects of regular semisimple type}
\label{sec:rst}

Similar to the $J^{(b)}[k]$ type of untwisted irregular punctures (defects) introduced in \cite{Wang:2015mra}, there is a distinguished class of twisted irregular punctures of the {\it regular semisimple} type. This is achieved when $T_k$ is a regular semisimple element
satisfying
% of $\mfj^k$ in \eqref{tg}:
\ie
(\sigma_g o) T_k=\omega^ k T_k. 
\label{tpr}
\fe
for $\omega=e^{2\pi i\over b_t}$ and a twisted torsion automorphism $\sigma_g o$ of $\mfj$. 

Restricted to the commutant Cartan subalgebra $\mfh$ of $T_k$,  $\sigma_g o$ induces an element in the twisted Weyl group
$W_t(\mfh)\equiv W(\mfh)  \rtimes{\rm Out}(\mfj)$. In particular, it corresponds to a twisted regular element of $W_t(\mfh)$ in Springer's classification  that generalizes the (untwisted) regular elements of $W(\mfh)$ which are associated to inner torsion automorphisms of $\mfj$ (for all simple Lie algebras) with regular eigenvectors \cite{springer1974regular}. 
Since we always work with semisimple elements of $\mfh$ in this paper, we will abuse the notation and use $\sigma_g o$ to denote both the  twisted torsion automorphism and the corresponding twisted regular element of $W_t(\mfh)$, similarly for $\sigma_g$ in the untwisted case.\footnote{For $\mfj=E_6$ or $E_7$ there are examples of a regular element (also twisted regular element for $E_6$) in the Weyl group that lift to two torsion automorphisms $\sigma_g,\sigma'_g$ or $\sigma_g o,\sigma'_g o$ for the twisted case (see Table~6.1, 6.2 and 6.6 in \cite{2012arXiv1205.0515E}). However we do not see any differences in the 4d theories engineered by punctures that are compatible with either $\sigma_g$ or $\sigma_g'$.} 
As in the untwisted case, each twisted regular element is  determined uniquely (up to conjugation) by a positive integer $d_t$ such that its regular eigenvector in $\mfh$ has eigenvalue $e^{2\pi i\over d_t}$. The order of the twisted regular element is  $d'_t=\lcm(d_t,|o|)$.\footnote{We emphasize that unlike in the untwisted case, the order does not determine the twisted regular element in general.}  In particular, the twisted regular element $\sigma_g o$ from \eqref{tpr} is associated with
\ie
\boxed{
~d_t={b_t\over \gcd(b_t,k)}.
~}
\fe
The twisted regular numbers $d_t$ were classified by Springer   \cite{springer1974regular} and summarized in Table~\ref{table:dt}.
%
%\noindent
%{$\pmb {{}^2A_{2N}}$}:
%(I) $d_t>2$ is an even divisor  of $2(2N+1)$,
%(II) $d_t$ is a divisor of $2N$.
%\\
%{  $\pmb { {}^2A_{2N-1}}$}:
%(I) $d_t>2$ is an even divisor  of $2(2N-1)$,
%(II) $d_t$ is a divisor of $2N$.
%\\
%{$\pmb { {}^2D_{N}}$}:
% (I)  $d_t>2$ is an even divisor  of $2N$
% (II) $d_t$ is a divisor of $2N-2$.	 
%\\
%{$\pmb { {}^2E_6}$}:
% (I)  $d_t=18$,
% (II) $d_t$ is a divisor of $12$,
% (III) $d_t$ is a divisor of $8$.
%\\
%{$\pmb { {}^3D_4}$}:
% (I)  $d_t=12$,
% (II) $d_t$ is a divisor of $6$.
  
\begin{table}[!htb]
	\begin{center}
		\begin{tabular}{ |c|c|c|c| }
			\hline
			\multirow{2}{*}{\parbox{2.0cm}{~~~~~${}^{|o|}\mfj $}}
    &  \multicolumn{2}{c|}{$d_t$}  
       \\ \cline{2-3}
    & $d_t \equiv 0 \,({\rm mod\,} |o|) $ is a divisor of  & $d_t $ is a divisor of  \\ \hline
			${}^2A_{2N}$ &  $4N+2$ & $2N$ \\ \hline
			${}^2A_{2N-1}$  & $4N-2$ & $2N$ \\ \hline
			${}^2D_N$    &$2N$ & $2N-2$  \\      \hline
			${}^2E_6$    & $18$ & $12,~8$\\   \hline
			${}^3D_4$     & $12$ & $6$ \\     \hline
		\end{tabular}
	\end{center}
	\caption{Twisted regular numbers $d_t$ for twisted Weyl groups.}
	\label{table:dt}
\end{table}

In particular the maximal values for $d_t$  correspond to the orders of twisted Coxeter elements and are called twisted Coxeter numbers $h_t$ of $\mfj$ in  \cite{springer1974regular} (see the second column Table~\ref{table:dt}).
In general, a twisted regular element induces a grading on the Cartan subalgebra
\ie
{\mf h}=\bigoplus_{m \in \mZ /d'_t\mZ} \mfh^m,\quad
\label{tgh}
\fe
such that $\mfh^{d'_t/d_t}$ contains a regular semisimple element. The corresponding twisted irregular Hitchin pole is specified by
\ie
\boxed{
b_t=d'_t,~\gcd(k,b_t)={d'_t/d_t},~{\rm and~} T_m\in \mfh^m~{\rm with~} T_k {\rm~regular~semisimple}.
}
\label{bkT}
\fe

We group these Hitchin poles into class I and II for ${}^2A_N, {}^2D_N, {}^3D_4$, and class I, II and III for ${}^2E_6$ according to the values of $d_t$ listed in the Table~\ref{table:dt}, with the understanding that their allowed divisors fall into the corresponding classes.\footnote{We emphasize here that the class I, II (and III for ${}^2 E_6$) of twisted defects are not always distinguishable.
If $d_t$ is a divisor of more than one entry in  Table~\ref{table:dt}, it defines a unique defect that appears in multiple classes.
} Below we give explicitly the Hitchin pole for $A_{2N-1},A_{2N}$ and $D_N$ when the associated twisted torsion automorphism $\sigma_g o$ has the maximal order in each class. 
 
For example, consider the case $\mf j=A_{2N-1}$. There are two  classes of twisted irregular punctures. Denoting the standard basis of $\mR^{2N}$ by $\{e_1,e_2,\dots,e_{2N}\}$ and the Cartan subalgebra of $A_{2N-1}$ by $\mR^{2N}/\{\sum_{i=1}^{2N}e_{i}\}$, the generator of ${\rm Out}(A_{2N-1})=\mZ_2$ is defined by
\ie
e_i \to -e_{2N+1-i}
\fe
while the Weyl group $W(A_{2N-1})$ acts by permutation on $e_i$.  
\subsubsection*{$A_{2N-1}$ Class I}
\ie
\Phi={ T \over z^{2+{k\over 4N-2}}}+\dots
\label{hp1}
\fe
for $\gcd(k,4N-2)=1$, where  
\ie
  T=\begin{pmatrix}
	0 & & & & &  
	\\
	& 1 & & & &  
	\\
	& & \omega^{2} & & &   
	\\
	&   &   &     \ddots & &  \\
	\\
	& & & & &   \omega^{ 4N-4 }  
\end{pmatrix},\quad \omega^{4N-2}=1
\fe
The required gauge transformation $\sigma_g$ corresponds to a permutation in the $\mZ_{2N-1}$  subgroup of $W(A_{2N-1})$ acting the lower right $2N-1$ diagonal entries of $T$. 
\subsubsection*{$A_{2N-1}$ Class II}
\ie
\Phi={ T \over z^{2+{k\over 2N}}}+\dots
\label{hp2}
\fe
for $\gcd(k,2N)=1$, where  
\ie
  T=\begin{pmatrix}
	1 & & & & &  
	\\
	& \omega & & & &  
	\\
	& & \omega^2  & & &   
	\\
	&   &   &     \ddots & &  \\
	\\
	& & & & &   \omega^{ 2N-1}  
\end{pmatrix},\quad \omega^{2N}=1
\fe
The required $\sigma_g$ now corresponds to a permutation in the $\mZ_{2N}$  subgroup of $W(A_{2N-1})$ .  
 
 The case $\mf j=A_{2N}$ is similar and again has two  classes of twisted irregular punctures. The generator of ${\rm Out}(A_{2N})=\mZ_2$ defined by\footnote{A special feature of the $\mZ_2$ outer-automorphism of $A_{2N}$ compared to the other cases in Table~\ref{table:outm} is that it does not fix any simple root, instead it exchanges the pair of simple roots $\A_{N}=e_N-e_{N+1}$ and $\A_{N+1}=e_{N+1}-e_{N+2}$. This has important consequences on the 4d theories engineered by such punctures which we explain later in the paper. }
\ie
e_i \to -e_{2N+2-i}.
\fe
while the Weyl group $W(A_{2N+1})$ acts by permutation on $e_i$ with $1\leq i\leq 2N+1$.  

\subsubsection*{$A_{2N}$ Class I}
\ie
\Phi={ T \over z^{2+{k\over 4N+2}}}+\dots
\label{hp1}
\fe
for $\gcd(k,4N+2)=1$, where  
\ie
 T=\begin{pmatrix}
	1 & & & & &  
	\\
	& \omega^2 & & & &  
	\\
	& & \omega^{4} & & &   
	\\
	&   &   &     \ddots & &  \\
	\\
	& & & & &   \omega^{ 4N }  
\end{pmatrix},\quad \omega^{4N+2}=1
\fe
 
\subsubsection*{$A_{2N}$ Class II}
\ie
\Phi={  T \over z^{2+{k\over 2N}}}+\dots
\label{hp2}
\fe
for $\gcd(k,2N)=1$, where  
\ie
  T=\begin{pmatrix}
	0 & & & & &  
	\\
	& 1 & & & &  
	\\
	& & \omega  & & &   
	\\
	&   &   &     \ddots & &  \\
	\\
	& & & & &   \omega^{ 2N-1}  
\end{pmatrix},\quad \omega^{2N}=1
\fe
 
For $\mf j=D_{N}$ there are two  classes of twisted irregular punctures. Identifying the Cartan subalgebra of $D_{N}$ with $\mR^N$ in the standard basis $\{e_1,e_2,\dots, e_N\}$, ${\rm Out}(D_{2N})=\mZ_2$ is generated (up to conjugation) by
\ie
 e_1\to - e_1,\quad e_i \to  e_i~{\rm for}~2\leq i\leq N
\fe
while the Weyl group $W(D_{2N})$ acts by permutation and even number of sign flips on $e_i$.  
  
\subsubsection*{$D_N$ Class I}
\ie
\Phi={ T \over z^{2+{k\over 2N}}}+\dots
\label{hp1}
\fe
for $\gcd(k,2N)=1$, where  
\ie
 T=\begin{pmatrix}
		0& 1 & & & & & &  \\
		-1& 0 & & & & & &  \\
		&   & 0 & \omega & & & &  \\
		&   & -\omega  & 0 & & & &  \\
		&   &   &   & & \ddots & &  \\
		&   &   &   & &  & 0 & \omega^{N-1}  \\
		&   &   &   & &  & -\omega^{N-1} & 0  \\		
	\end{pmatrix},\quad \omega^{2N}=1
\fe
 
\subsubsection*{$D_N$ Class II}
\ie
\Phi={  T \over z^{2+{k\over 2N-2}}}+\dots
\label{hp2}
\fe
for $\gcd(k,2N-2)=1$, where  
\ie
  T=\begin{pmatrix}
		0& 0 & & & & & &  \\
		0& 0 & & & & & &  \\
		&   & 0 & 1 & & & &  \\
		&   & -1 & 0 & & & &  \\
		&   &   &   & & \ddots & &  \\
		&   &   &   & &  & 0 & \omega^{N-1}  \\
		&   &   &   & &  & -\omega^{N-1} & 0  \\
		\end{pmatrix},
	\quad \omega^{2N-2}=1.
\fe

\subsection{Physical parameters from the punctures}

The defining data of the punctures can be identified with the parameters of the resulting 4d theories. In particular, for SCFTs, we are interested in the masses for flavor symmetries and exactly marginal couplings. Of course one can enumerate such parameters in the SW (spectral) curve. Here we describe how these data can be extracted directly from the punctures. 

The grading \eqref{tg} induces a natural conjugation action of a reductive Lie group $J_0$ associated to $\mfj^0$ on $\mfj^\ell$. Therefore $\mfj^\ell$ are $J_0$-modules of finite ranks. Furthermore, the $U(1)_R$ symmetry associated with the singularity 
acts by 
\ie
U(1)_R \subset SO(2)_{45} \times U(1)_z:~ \Phi\to e^{i\A}\Phi,\quad z \to e^{i \A {b_t\over k+b_t}} z.
\fe
such that $T_k \in \mfj^{d'_t/d_t}$ has zero $U(1)_R$ charge. In general $T_\ell$ has $U(1)_R$ charge
\ie
q_R[T_\ell]={k-\ell\over k+b_t}
\fe
Hence $T^{-b_t} \in \mfj^0$ has $U(1)_R$ charge $q_{R}=1$. From $\cN=2$ superconformal symmetry, we deduce that $T^{-b_t}$ contains the mass parameters of the theory whereas $T_k$ is associated to exactly marginal couplings.  

The maximal number of exactly marginal couplings is determined by the rank of $\mfj^{d'_t/d_t}$ as a $J_0$-module to be
\ie
n_{\rm marg}=\rank(\mfj^{d'_t/ d_t}|J^0)-1= \dim (\mfh^{d'_t/ d_t})-1.
\fe
Note that we have fixed the redundancy due to conjugation by $J_0$ as well as rescaling of the $z$ coordinate.

The maximal number of mass parameters is captured by the dimension of the intersection between the centralizer of the semisimple part of $\mfj^{d'_t/ d_t}$ and $\mfj^0$
\ie
n_{\rm mass}=\dim (C(\mfj^{d'_t/ d_t}_s)\cap \mfj^0)= \dim (\mfh^0).
\fe
Both $n_{\rm marg}$ and  $n_{\rm mass}$ can be extracted from the grading \eqref{tg} of $\mfh$ described in \cite{springer1974regular}. We summarize the results in Table~\ref{table:aedata}-\ref{table:d4data}. 
 
\begin{table}[!htb]
	\begin{center}
		\begin{tabular}{ |c|c|c| }
			\hline
		$d_t   $ & Number of marginal couplings   &  Number of  mass parameters
		   \\ \hline
%			$d_t=2$  & $2N-1$ &   $2(2N+1)/d_t-1$ \\ 
 $d_t=1$ &   $N-1$ & $N$ \\  
		$d_t \in  2\mZ+1$,~$d_t|N$ & $N/d_t-1$ &   $N/d_t$ \\  
			$d_t \in 4\mZ$,~$d_t|2N$ & $2N/d_t-1$ &   $2N/d_t$ \\ 
			$d_t\in 4\mZ+2$,~$d_t|2(2N) $ & $4N/d_t-1$ &   $0$ \\  
	$d_t>2 \in 4\mZ+2$,~$d_t|2(2N+1) $ & $2(2N+1)/d_t-1$ &   $0$ \\    	 \hline
				\end{tabular}
	\end{center}
	\caption{Marginal couplings and mass parameters from twisted $A_{2N}$ punctures}
	\label{table:aedata}
\end{table}

\begin{table}[!htb]
	\begin{center}
	\begin{tabular}{ |c|c|c| }
		\hline
		$d_t$ &  Number of marginal couplings   &  Number of mass parameters
		\\ \hline
%			$d_t=2$  & $2N-1$ &   $2(2N+1)/d_t-1$ \\ 
$d_t=1$ &   $N-1$ & $N$ \\  
			$d_t>1 \in  2\mZ+1$,~$d_t|N$ & $N/d_t-1$ &   $N/d_t$ \\  
		$d_t \in 4\mZ$,~$d_t|2N$ & $2N/d_t-1$ &   $2N/d_t$ \\ 
			$d_t>2 \in 4\mZ+2$,~$d_t|2(2N) $ & $4N/d_t-1$ &   $0 $ \\  
			$d_t \in 4\mZ+2$,~$d_t|2(2N-1) $ & $2(2N-1)/d_t-1$ &   $0$ \\   	 \hline
	\end{tabular}
\end{center}
	\caption{Marginal couplings and mass parameters from twisted $A_{2N-1}$ punctures}
	\label{table:aodata}
\end{table}

\begin{table}[!htb]
	\begin{center}
		\begin{tabular}{ |c|c|c| }
			\hline
			$d_t$ & Number of marginal couplings   &  Number of mass parameters
			\\ \hline
		$d_t=1$ &   $N-2$ & $N-1$ \\  
			$d_t \in  2\mZ+1$,~$d_t|(N-1)$ & $(N-1)/d_t-1$ &   $(N-1)/d_t$ \\  
			$d_t \in 2\mZ$,~$(2(N-1))/d_t\in 2\mZ+1$ & $2(N-1)/d_t$ &   0 \\ 
			$d_t \in 2\mZ$,~$(2(N-1))/d_t\in 2\mZ$  & $2(N-1)/d_t-1$ &   1 \\ 
			$d_t>2 \in 2\mZ $,~$2N/d_t\in 2\mZ+1 $ & $2N/d_t-1$ &   $0$ \\    \hline
				\end{tabular}
	\end{center}
	\caption{Marginal couplings and mass parameters from twisted $D_{N}$ punctures}
	\label{table:ddata}
\end{table}

\begin{table}[!htb]
	\begin{center}
		\begin{tabular}{ |c|c|c| }
			\hline
			$d_t$ &Number of marginal couplings   &  Number of mass parameters
			\\ \hline
			1 & 3 &   4\\  
		       	2 & 5 &   0\\  
				3 & 1 &   0\\  
				4 & 1 &   0\\   
				6 & 2 &   0\\ 
				8 & 0 &   1\\ 
				12 & 0 &   0\\ 
				18 & 0 &   0\\ 				
				   \hline
		\end{tabular}
	\end{center}
	\caption{Marginal couplings and mass parameters from twisted $E_6$ punctures}
	\label{table:e6data}
\end{table}

\begin{table}[!htb]
	\begin{center}
		\begin{tabular}{ |c|c|c| }
			\hline
			$d_t$ & Number of marginal couplings   &  Number of mass parameters
			\\ \hline
				1  & 1 &  2\\  
			2 & 1 &  2\\  
			3 & 1 &   0\\  
			6 & 1 &   0\\   
			12 & 0 &   0\\ 
					\hline
		\end{tabular}
	\end{center}
	\caption{Marginal couplings and mass parameters from twisted $D_4$ punctures}
	\label{table:d4data}
\end{table}

%%%%%%%%%%%%%%%%%%%%%%
%
\newpage

\section{Twisted theories and central charges}
\label{sec:tt}

Given the classification of the twisted irregular defects in the previous section, we now use them to construct 4d $\cN=2$ SCFTs and study properties of the resulting theories. In particular, we determine their flavor symmetry and Coulomb branch spectrum, and propose conjectured formulae for the flavor and conformal central charges. We later offer nontrivial checks for these conjectures.

\subsection{Classification of theories from twisted irregular defects}

\begin{figure}[htb]
	\centering
	%	\begin{minipage}{0.3\textwidth}
	%		\includegraphics[width=.8\textwidth]{utsp.pdf}
	%	\end{minipage}%
	\begin{minipage}{0.3\textwidth}
		\includegraphics[width=.8\textwidth]{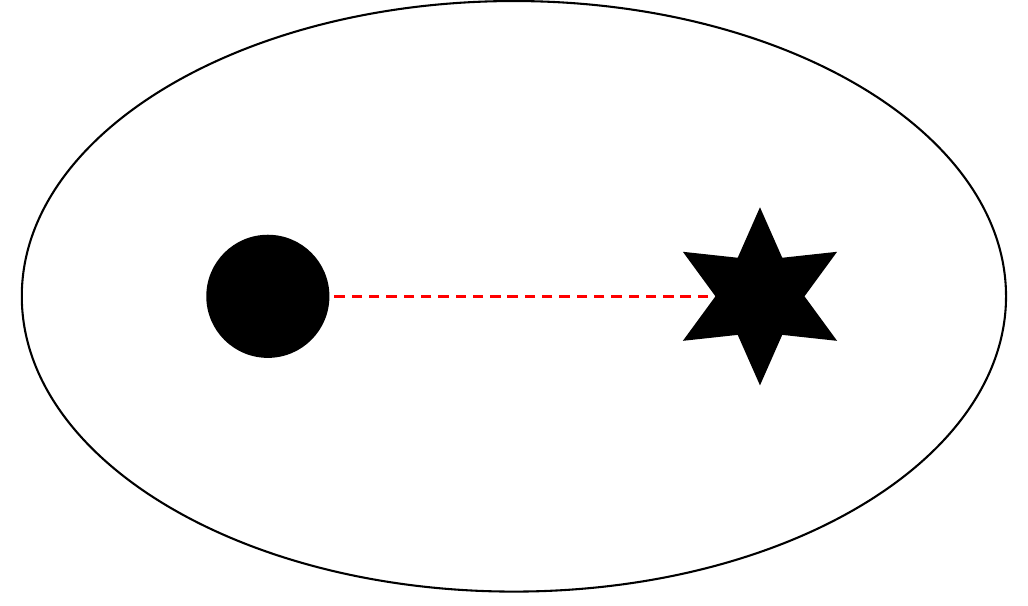}
	\end{minipage}%
	\caption{The class S setup for twisted theories: one twisted irregular defect (star) and one twisted regular defect (dot) on a sphere.}
	\label{fig:tsp}
\end{figure}

We are interested in 4d $\cN=2$ SCFTs engineered by compactifiying six dimensional $(2,0)$ SCFT of ADE type on twice-punctured $\mP^1$ with outer-automorphism twist. The twist line connects one twisted irregular singularity and one twisted regular singularity on $\mP^1$. We will refer to such 4d theories as twisted theories (see Figure~\ref{fig:tsp}).
One common feature of the twisted theories is their non-simply-laced flavor groups $G$ (see Table~\ref{table:outm}) coming from the twisted regular punctures. 

 As we have reviewed in the last section, the classification of twisted irregular punctures is reduced to that of torsion outer-automorphisms of the Lie algebra $\mfj$ and the  associated gradings \eqref{tg}. In this section, we focus on the regular semisimple type (see Section~\ref{sec:rst}). The torsion outer-automorphism in this case corresponds to a regular element of the twisted Weyl group $W_t(\mfh)$ and induces a grading \eqref{tgh} on the Cartan subalgebra $\mfh$,  labelled by  $d_t$ in Table~\ref{table:dt}.

The corresponding twisted irregular singularity of the Higgs field takes the following forms  
\ie
& \Phi_z={T\over z^{2+{2k'+1\over b_t}}}+\ldots,~~{}^{|o|}\mfj={}^2 A_N, {}^2 D_N, {}^2 E_6 ~{\rm and}~b_t=h_t 
\\
&  \Phi_z={T\over z^{2+{3k'\pm1\over b_t}}}+\ldots,~~{}^{|o|}\mfj={}^3 D_4~{\rm and}~b_t=h_t 
\\
& \Phi_z={T\over z^{2+{k'\over b_t}}}+\ldots,~~b_t\neq h_t. 
\label{thptogether}
\fe
Here $T$ is regular semisimple, $b_t$ takes the values of $d_t'$ as in \eqref{bkT} and $k'$ is an arbitrary integer such that the leading  pole order is larger than one. Recall that $h_t$ denotes the twisted Coxeter number.  

%\ie
%aaa
%\label{table:SW}
%\fe
For simplicity we take the regular twisted puncture to be maximal.  Some physical properties of the resulting twisted theories can be extracted as follows:
\begin{itemize}
\item  The SW curve at the conformal point takes the forms listed in Table~\ref{table:SW}. Note that 
now the $z$ variable admits fractional powers.
 
%
%
%\begin{table}[!htb]
%\begin{center}
%\begin{tabular}{|c|c|c|c|}
%\hline
%Flavor group $G$ & $b_t$ & SW geometry at SCFT point & $\Delta[z]$ \\ \hline
% $C_N^\text{{anom}}$ & $4N+2$ &$x_1^2+x_2^2+x^{2N+1}+z^{k+{1\over2}}=0$ & ${4N+2\over 4N+2k+3}$  \\ \hline
% ~& $2N$ & $x_1^2+x_2^2+x^{2N+1}+xz^k=0$ & ${2N\over k+2N}$ \\ \hline
%  $B_N$& $4N-2$ & $x_1^2+x_2^2+x^{2N}+xz^{k+{1\over2}}=0$ & ${4N-2\over 4N+2k-1}$ \\ \hline
%~ & $2N$ &$x_1^2+x_2^2+x^{2N}+z^{k}=0$ & ${2N\over 2N+k}$  \\ \hline
% $C_N^\text{\sout{anom}}$& $2N+2$ & $x_1^2+x_2^{N}+x_2x_3^2+x_3z^{k+{1\over2}}=0$ & ${2N+2\over 2k +2N+3}$ \\ \hline
%~ & $2N$ &$x_1^2+x_2^{N}+x_2x_3^2+z^{k}=0$ & ${2N\over k+2N}$  \\ \hline
% $G_2$ & $12$ &$x_1^2+x_2^{3}+x_2x_3^2+x_3z^{k\pm {1\over3}}=0$ & ${12\over 12+3k\pm1}$  \\ \hline
%  ~& $6$ &$x_1^2+x_2^{3}+x_2x_3^2+z^{k}=0$ & ${6\over 6+k}$  \\ \hline
%$F_4$& $18$ &$x_1^2+x_2^{3}+x_3^4+x_3z^{k+{1\over2}}=0$ & ${18\over 18+2k+1}$  \\ \hline
%  ~& $12$ &$x_1^2+x_2^{3}+x_3^4+z^{k}=0$ & ${12\over 12+k}$  \\ \hline
%~ & $8$ &$x_1^2+x_2^{3}+x_3^4+x_2z^{k}=0$ & ${8\over 12+k}$  \\ \hline
%\end{tabular}
%\caption{Seiberg-Witten geometry at the SCFT point.}
%\label{table:SW}
%\end{center}
%\end{table}

\begin{table}[htb]
\begin{center}
\begin{tabular}{|c|c|c|}
\hline
$\mfj$&Degrees of the Casimirs $d_i$ & ~ Transformation under $o$ \\ \hline
$A_{N}$&$2,3,\ldots, N$ & $\phi_{d_i}\to (-1)^{d_i}\phi_{d_i}$ \\ \hline
 & &
\\[-1em]
$D_{N}$&$2,4,\ldots, 2N-2,N$ & $\tilde{\phi}_{N}\to-\tilde{\phi}_{N}$ \\ \hline
 & &
\\[-1em]
$D_{4}$&$2,4,4$ & $\phi_4\to e^{2\pi i\over 3}\phi_4,~\tilde{\phi}_4\to e^{4\pi i\over 3}\tilde{\phi}_4$ \\ \hline
 & &
\\[-1em]
$E_6$ & $2,5,6,8,9,12$ & $\phi_{d_i}=(-1)^{d_i}\phi_{d_i}$ \\ \hline
\end{tabular}
\caption{Casimirs and their transformations under outer-automorphisms}
\label{table:transf}
\end{center}
\end{table}

\item The regular twisted puncture gives rise to non-Abelian flavor symmetry $G$ (see Table~\ref{table:outm}). There are two cases with $C_N$ flavor symmetry from $\mZ_2$ twisted defects in $A_{2N}$ and $D_{N+1}$ $(2,0)$ theories. We label them by $C^{\rm anom}_N$ and $C^\text{\sout{anom}}_N$ respectively.\footnote{We thank Yuji Tachikawa for suggesting these names.}
They differ by Witten's global anomaly for $USp(2N)$ \cite{Tachikawa:2018rgw}: the former carries a nontrivial anomaly whereas the latter is non-anomalous.

\item The number of mass parameters for $U(1)$ flavor symmetries associated to the twisted irregular  puncture comes from $\Delta=1$ parameters in the Hitchin poles, as summarized in  Table~\ref{table:aedata}-\ref{table:d4data}.

\item The dimension of the conformal manifold can be extracted from $\Delta=0$ parameters for the Hitchin poles, as summarized in  Table~\ref{table:aedata}-\ref{table:d4data}.

\item The full Coulomb branch spectrum can be found by expanding the degree $d_i$ differentials $\phi_{d_i}(z)$ in $z$,
where $d_i$ labels the degrees of the Casimirs of the parent ADE theory, see Table~\ref{table:transf} for these numbers. 
The novelty here is that some of the differentials are no longer holomorphic and they often have a Laurent 
expansion with half integral (or plus-minus one third) powers of $z$, according to their transformation rule under the outer-automorphism in Table~\ref{table:transf}.
The spectrum of half-BPS Coulomb branch chiral primaries is then summarized in Table~\ref{table:coulomb}.

\begin{table}[h]
\begin{center}
\begin{tabular}{|c|l|}
\hline
Flavor group $G$ & Coulomb branch spectrum $\Delta$ \\ \hline
 $C_N^\text{{anom}}$ & $d_i-{2j+1\over 2}\Delta[z]>1,~~d_i=3,5,\ldots, 2N +1,~j\geq 0$ \\ 
 ~&$d_i-j\Delta[z]>1,~~d_i=2,4,\ldots, 2N,~j\geq 1$ \\ \hline
 $B_N$& $d_i-{2j+1\over 2}\Delta[z]>1,~~d_i=3,5,\ldots, 2N-1,~j\geq 0$ \\ 
~&$d_i-j\Delta[z]>1,~~d_i=3,5,\ldots, 2N,~j\geq 1$ \\  \hline
 $C_N^\text{\sout{anom}}$ & $d_i-{2j+1\over 2}\Delta[z]>1,~~d_i=N+1,~j=0,\ldots$ \\
 ~& $d_i-{j}\Delta[z]>1,~~d_i=2,\ldots,2N,~j=1,\ldots$ \\ \hline
 $G_2$ & $d_i-{3j+1\over 3}\Delta[z]>1,~~d_i=4,~j\geq 0$ \\
~& $d_i-{3j+2\over 3}\Delta[z]>1,~~d_i=4,~j\geq 0$ \\
 ~& $d_i-{j}\Delta[z]>1,~~d_i=2,~j\geq 1$ \\ \hline
 $F_4$ & $d_i-{2j+1\over 2}\Delta[z]>1,~~d_i=5,9,~j\geq 0$ \\
 ~& $d_i-{j}\Delta[z]>1,~~d_i=2,6,8,12,~j\geq 1$ \\ \hline
\end{tabular}
\caption{Coulomb branch spectrum of the twisted theories from a twisted irregular puncture of the regular-semisimple type and a maximal twisted regular puncture. The Coulomb branch chiral primaries are constrained to have $\Delta>1$ by unitarity.}
\label{table:coulomb}
\end{center}
\end{table}

\begin{table}[h]
\begin{center}
\begin{tabular}{|c|c|c|c|c|c|}
\hline
~&dimension & $h$ & $h^{\vee}$&  $r^\vee$ & $n$ \\ \hline
$A_{N-1}$&$N^2-1$& $N$ & $N$&1  &1\\ \hline
$B_N$ & $(2N+1)N$ & $2N$& $2N-1$&2 &2\\ \hline
$C_N^\text{{anom}}$ &  $(2N+1)N$&  $2N$ &$N+1$&2&4 \\ \hline
$C_N^\text{\sout{anom}}$ &  $(2N+1)N$&  $2N$ &$N+1$&2 &2\\ \hline
 $D_N$ & $N(2N-1)$ & $2N-2$ & $2N-2$&1 &1\\ \hline
$E_6$& 78 & 12 & 12 &1&1\\ \hline
$E_7$& 133 & 18 & 18 &1&1\\ \hline
$E_8$& 248 & 30 & 30 &1&1\\ \hline
$F_4$& 52 & 12 & 9 &2&2\\ \hline
$G_2$& 14 & 6 & 4&3&3 \\ \hline

\end{tabular}
\end{center}
\caption{Some useful Lie algebra data. $h$ is the Coxeter number and $h^{\vee}$ is the dual Coxeter number. $r^\vee$ is the lacety of the Lie algebra and $n$ is equal to $r^\vee$  except for $C_N^\text{{anom}}$. }
\label{table:lie}
\end{table}

\item We present conjectures for the flavor and conformal central charges with nontrivial evidences in the next section.

\end{itemize}

 \subsection{Flavor and conformal central charges}

 Before we state the conjectural formulae for the flavor and conformal central charges, let us provide some motivations for how they come about. The  important observation is, in all previously known class S constructions that involve maximal regular punctures,  it appears that the flavor central charge is determined by certain maximal scaling dimension among the CB spectrum contributed by the maximal puncture:\footnote{Note that our normalization of the flavor central charges is related to the one in \cite{Chacaltana:2012zy,Chacaltana:2014nya} by $k_G^{\rm ours}={1\over 2}k_G^{\rm theirs }$.}
\ie
k_{G}=& \Delta_{\rm max}, &\quad G=A_N,D_N,E_N
\\
k_{G}=& \widetilde\Delta_{\rm max},&\quad G=B_N,C_N^\text{\sout{anom}},\,F_4,\,G_2
\\
k_{G}=&{1\over 2}\widetilde\Delta_{\rm max}+{1\over 2},&\quad G=C_N^\text{{anom}}
\label{fcobs}
\fe
Here $\Delta_{\rm max}$ denotes the maximal scaling dimension contributed by the maximal regular puncture which determines the flavor central charge in the untwisted theories. In twisted theories, empirical evidence suggests that one should instead take the maximal scaling dimension $\widetilde \Delta_{\rm max}$ from the {\it twisted} differentials (i.e. a differential that transforms non-trivially under the twist, see Table~\ref{table:transf}) at  the maximal puncture. 
 The $\mZ_2$ twisted maximal punctures of $A_{2N}$ type requires special attention.\footnote{Note that the ${1\over 2}$ shift in the third equation of \eqref{fcobs} is exactly the contribution from a half-hyper multiplet in the fundamental representation of $USp(2N)$ which would also saturate Witten's global anomaly for these punctures \cite{Tachikawa:2018rgw}. 
Thus it's tempting to say that this ${1\over 2}$ contribution comes from the \textit{minimal} boundary modes of $\mZ_2$ symmetry defect.
The other ${1\over 2}$ factor in \eqref{fcobs} should be related to the fact that the Langlands dual $SO(2N+1)$ of the $USp(2N)$ has an index of embedding equal to one in $D_{N+1}$ but {two} in $A_{2N}$. A better understanding of the anomaly inflow from 6d along the lines of \cite{Bah:2018gwc} should give a rigorous argument for \eqref{fcobs}.
 } For example, in a $\mZ_2$ twisted $A_{2N}$ type class S construction without irregular punctures, the maximal scaling dimension contributed by the twisted differentials at the maximal twisted regular singularity is $2N+1$ from $\phi_{2N+1}$, and the $USp(2N)$ flavor central charge computed in \cite{Chacaltana:2014nya} is equal to (in our normalization) $h^\vee(USp(2N))=N+1$ in agreement with \eqref{fcobs}. The introduction of irregular punctures into the setup modifies the $U(1)_R$ symmetry of the 4d theory, but we expect \eqref{fcobs} continues to hold. The formula for $2a-c$ is given in \cite{Shapere:2008zf} while our new formula for $c$ has a close relation to the 2d chiral algebra which we will explain in Section~\ref{sec:voa}.\footnote{The essential statement here is that the 2d chiral algebra 
 contains the affine Kac-Moody algebras associated to both the simple and $U(1)$ factors of the flavor symmetry,  and the 2d stress-tensor is given by the Sugawara construction.}

\begin{conj}[Flavor central charge $k_G$]
The  central charge of the flavor symmetry $G$ of a  twisted theory defined by a twisted irregular defect of the regular semisimple type in \eqref{thptogether} and a maximal twisted regular defect takes the following form:
\ie
\boxed{
	~~ k_G=h^{\vee}-{1\over n}{b\over l+b},~~~\begin{cases}
		l=2k'+1~~~b_t=h_t,~G\neq G_2 \\
		l=3k'\pm1~~~b_t=h_t,~G= G_2 \\
		l=k'~~~~~~~~~ b_t\neq h_t.
	\end{cases}
}
\label{fla}
\fe 
The Lie algebra data $h^\vee$ and $n$ are listed in Table~\ref{table:lie}.
\end{conj}
 
\begin{conj}[Conformal central charges $a$ and $c$] 
	The conformal central charges of a general twisted theory are determined by
	\ie
\boxed{
2a-c={1\over 4}\sum (2\Delta[u_i]-1),~~c={1\over 12}{k_G \dim(G) \over -k_G+h^{\vee}}-{f\over 12}
\label{conjcc}
}
\fe
Here $f$ is the number of mass parameters contributed by the irregular singularity, $k_G$ denotes the flavor symmetry  central charge
listed in \eqref{fla}, and $\Delta[u_i]$ are the Coulomb branch scaling dimensions listed in Table~\ref{table:coulomb}. 
\end{conj}

Let us provide some evidences for the above conjectures by considering twisted theories defined with integral order Hitchin poles, in which case the irregular singularity takes the   form
\ie
\Phi_z={T\over z^{2+k}}+\ldots
\label{integralpole}
\fe
where $T$ is regular semisimple and the associated grading of $\mfh$ corresponds to $d_t=1$. We will use following formula
\ie	
2a-c={1\over 4}\sum_{i}(2[u_i]-1),~~~a-c=-{\dim_\mH{\rm Higgs} \over 24}
\label{higgs}
\fe
to compute their central charges. The second equation above is known to hold for the untwisted irregular puncture defined by integral order Hitchin poles \cite{Xie:2012hs}  and we assume that it is also valid for the current situation.\footnote{The equation that relates $a-c$ to the Higgs branch dimension is a  simple consequence of anomaly matching  \cite{Shimizu:2017kzs}. It should hold whenever the Higgs branch is {\it pure}:  the low energy theory on the Higgs branch is described just by hypermultiplets. It would be interesting to understand how it gets modified for general Argyres-Douglas type SCFTs.} 

Here we take the regular twisted puncture to be maximal in which case the Higgs branch dimension is \cite{Chacaltana:2012zy,Wang:2015mra}
\ie
\dim_\mH{\rm Higgs}={1\over 2}(\dim(G)-\rank(G))+\rank(G).
\label{hb}
\fe
The last term above comes from the twisted irregular puncture.

On the other hand, since the irregular singularity contributes $f=\rank(G)$ mass parameters (see top rows of Table~\ref{table:aedata}-\ref{table:d4data}), 
the conjectural formula \eqref{conjcc} for central charge $c$ takes the following form 
\ie
c={1\over 12}{k_G  \dim(G) \over -k_G+h^{\vee}}-{1\over 12}\rank(G).
\label{4d2dca}
\fe
Here the conjectured flavor central  charge \eqref{fla} takes the form 
\ie
k_G=h^{\vee}-{1\over n}{1\over k+1}.
\label{4dkg}
\fe
We have verified that that the two formulae \eqref{4d2dca} and \eqref{higgs} give the same answers. Below we give some details for two instances of such checks for illustration. 
\setcounter{theorem}{0}
\theoremstyle{definition}
\begin{eg}
	\theoremstyle{definition}
 Let's consider the $C^{\text{\sout{anom}}}_{N-1}$ theory which is constructed by $\mZ_2$ twist of the $D_N$  $(2,0)$ theory. The Hitchin pole \eqref{integralpole} fixes the $U(1)_R$ charge (hence scaling dimension) of $z$,
\ie
\Delta[z]={1\over k+1}
\fe
and the Coulomb branch spectrum can be enumerated using Table~\ref{table:coulomb}  
\ie
\Delta=\{2i-{j\over k+1}>1 | i=1,2,\dots,N-1, j\geq 1\}\bigsqcup\{N-{2j+1\over 2(k+1)}>1 | j\geq 0\}.
\fe
Therefore
\ie
& 2a-c={1\over 4}\sum_{i}(2\Delta[u_i]-1)=\frac{1}{12} (N-1) N (4 k N-2 k+4 N-5).
\fe
Along with the Higgs branch dimension from \eqref{hb}
\ie
& \dim_\mH{\rm Higgs}=  (N-1) N 
\fe
we obtain from \eqref{higgs}
\begin{align}
& a= \frac{1}{24} (N-1) N (8 k N-4 k+8 N-9),~c=\frac{1}{6} (N-1) N (2 k N-k+2 N-2)
\end{align}
which is in agreement with \eqref{4d2dca} where the flavor central charge is determined by \eqref{4dkg} to be
\ie
k_{C_{N-1}}=N-{1\over 2(k+1)}.
\fe.
\end{eg}

\begin{eg} Similarly let us consider the $F_4$ theory which is engineered by $\mZ_2$ twist of the $E_6$ $(2,0)$ theory with irregular punctures. Once again
\ie
\Delta[z]={1\over k+1}
\fe
and the Coulomb branch spectrum is
\ie
\Delta=\{d-{j\over k+1}>1 | d=2,6,8,12~{\rm and}~j\geq 1\}\bigsqcup\{d-{2j+1\over 2(k+1)}>1 | d=5,9~{\rm and}~j\geq 0\}
\fe
 from Table~\ref{table:coulomb}, which gives
\ie
& 2a-c={1\over 4}\sum_{i}(2[u_i]-1)=78 k+71. 
\fe
Next the Higgs branch dimension follows from \eqref{hb}
\ie
&  \dim_\mH{\rm Higgs}=28.
\fe
We thus obtain from \eqref{higgs}
\ie
& a= 78 k+\frac{433}{6},~c=78 k+\frac{220}{3}
\fe
which is in agreement with \eqref{4d2dca} where 
\ie
k_{F_4}=9-{1\over 2(k+1)}.
\fe
from \eqref{4dkg}. 
\end{eg}

\subsection{Twisted theories with Lagrangians}
It turns out that many sub-families of the theories engineered from twisted irregular defects actually have Lagrangian descriptions.\footnote{In this paper, we implicitly assume that the SW geometry together with all of its $\cN=2$ deformations fixes the $\cN=2$ SCFT uniquely. To our best knowledge there is no counter-example but it would be interesting to prove this rigorously.} Since the Coulomb branch spectrum for Lagrangian theories have integral scaling dimensions, a necessary condition is   $\Delta[z]\equiv 0\,({\rm mod}~2)$ for $A_{n},D_n,E_6$ theories with $\mZ_2$ twist, and $\Delta[z]\equiv 0\,({\rm mod}~3)$ for $D_4$ with $\mZ_3$ twist. This can be achieved by choosing $k$ appropriately with respect to $b_t$ in the Hitchin pole \eqref{thptogether}.

Since such theories have a weakly coupled frame, we can use the formulae
\ie
a={5n_v\over 24}+{n_h\over 24},~~~c={n_v\over 6}+{n_h\over 12}
\label{weakcc}
\fe
to compute the conformal central charges.  Here $n_v$ and $n_h$ count the number of vector and hypermultiplets in the quiver gauge theory description.
Similarly the central charges associated to the $G_{\rm flavor}$ flavor symmetry of hypermultiplets can be obtained straightforwardly from identifying the embedding $G_{\rm flavor}\times G_{\rm gauge} \subset USp(2n_h)$. This allows us to verify the conjectured formulae \eqref{fla} and \eqref{conjcc}.
 
 As we will see, often times the Lagrangian description only makes manifest a subgroup of the full flavor symmetry which is realized in our description by the single regular puncture.

\begin{eg}
\label{eg:3}
Let's take $C^\text{\sout{anom}}_N$ which comes from the $\mZ_2$ twist of $D_{N+1}$ $(2,0)$ theory. The twisted irregular puncture is specified by 
\begin{equation}
b_t=2N,\quad k=-N
\end{equation}
so that we have the scaling dimension $\Delta[z]=2$. The Coulomb branch spectrum for this theory (using Table \ref{table:SW} and \ref{table:coulomb}) is 
\ie
& \Delta=\bigsqcup_{j=1}^{N-1}\{2i|1\leq i\leq j\}\bigsqcup\{N-2i>1|| i\geq 0\}
%\\
%&  \text{Coulomb spectrum}: (2) (2,4) (2,4,6)\ldots, (2,4,\ldots, 2N-2), (N,N-2,\ldots, 2) \nonumber\\
\fe
The central charges from \eqref{conjcc} and \eqref{fla} are
\ie
N~{\rm even}:~& k_{USp(2N)}=N,\quad
 a={1\over 24}N(1+N+4N^2),\quad c={1\over12}N^2(1+2N).
 \\
 N~{\rm odd}:~& k_{USp(2N)}=N,\quad
 a={1\over 24}(-4+N+N^2+4N^3),\quad c={1\over12}(-1+N^2+2N^3).
 \label{kaceg3}
\fe
Note that from Table~\ref{table:ddata}, the $N$ even case has $N$ marginal couplings and no mass parameters from the irregular puncture whereas the $N$ odd case has  $N-1$ marginal couplings and an extra mass parameter. This results in the different expressions for the conformal central charges above.

On the other hand, this theory has a quiver gauge theory description by
\begin{center}
\begin{tabular}{ccc}
%\xymatrixrowsep{.1in}
\xymatrixcolsep{.1in}
	\xymatrix{
		\boxed{USp(2N)}  \ar@{-}[r] &SO(2N) \ar@{-}[r] & USp(2N-4) \ar@{-}[r] &SO(2N-4) \ar@{-}[r] 
		 \ar@{-}[r] & \dots \ar@{-}[r]
		 & SO(8) \ar@{-}[r] &USp(4)  \ar@{-}[r] & SO(4)	}
\end{tabular}
\end{center}
for $N$ even, and 
\begin{center}
\begin{tabular}{ccc}
%\xymatrixrowsep{.1in}
\xymatrixcolsep{.1in}
	\xymatrix{
		\boxed{USp(2N)}  \ar@{-}[r] &SO(2N) \ar@{-}[r] & USp(2N-4) \ar@{-}[r] &SO(2N-4) \ar@{-}[r] 
		 \ar@{-}[r] & \dots \ar@{-}[r]
		 & SO(6) \ar@{-}[r] &USp(2)  \ar@{-}[r] & \boxed{SO(2)}	}
\end{tabular}
\end{center}
for $N$ odd. It is easy to see that the quiver is balanced thus all vector multiplets are conformally gauged.

The boxed nodes label the flavor symmetry of hypermultiplets. Here the symmetry is $USp(2N)$ with central charge
\ie
k_{USp(2N)}=N
\fe
as provided by $2N$ hypermultiplets in the fundamental representation of $USp(2N)$.  We can also count  the number of hyper and vector multiplets from the quiver.
\ie
N~{\rm even}:~&n_h={1\over3}N(-2+3N+2N^2),~~n_v={1\over3}(N+2N^3)
\\
N~{\rm odd}:~&n_h=1+{2\over3}N(-2+N+2N^2),~~n_v={1\over3}(-3+N+2N^3)
\fe
which gives the central charges by \eqref{weakcc}
\ie
N~{\rm even}:~&a={1\over 24}N(1+N+4N^2),~~c={1\over12}N^2(1+2N)
\\
N~{\rm odd}:~&a={1\over 24}(-4+N+N^2+4N^3),~~ c={1\over12}(-1+N^2+2N^3)
\fe
Both the flavor and conformal central charges from the Lagrangian are in agreement with \eqref{kaceg3} obtained from our conjectured formulae \eqref{fla} and \eqref{conjcc}.
\end{eg}

\begin{eg}
 
  Let's take $G=C_N^\text{{anom}}$ which comes from the $\mZ_2$ twist of $A_{2N}$ $(2,0)$ theory. 
  The twisted irregular puncture is specified by 
  \begin{equation}
  b=4N+2,~~k=-2N-1
  \end{equation}
so that $\Delta[z]=2$. As before, the Coulomb branch spectrum (from Table~\ref{table:coulomb}) and central charges (from \eqref{fla} and \eqref{conjcc}) are
\ie
& \Delta=\bigsqcup_{j=1}^{N-1}\{2i|1\leq i\leq j\} \bigsqcup_{j=1}^{N-1}\{2i|1\leq i\leq j\}\bigsqcup\{ 2i || 1\leq i\leq N\}
%\\
%\Delta=&\{2N,~2N-2,~\dots,2,
% 2N-2,~2N-4,~\dots,2,
% 2N-2,~2N-4,~\dots,2,
% \dots,
% 2
% 2
%\}\\
\\
&k_G=N+{1\over2} ,
\quad
a={N(8N^2+7N+3)\over 24} ,\quad c={N(1+2N)^2\over 12}.
\label{kaceg4}
\fe
In particular, the irregular singularity contributes no additional mass parameters and the theory has $2N-1$ marginal couplings   (see Table~\ref{table:aedata}).   The theory has a Lagrangian description by
\begin{center}
\begin{tabular}{ccc}
%\xymatrixrowsep{.1in}
\xymatrixcolsep{.1in}
	\xymatrix{
		\boxed{USp(2N)}  \ar@{-}[r] &SO(2N+1) \ar@{-}[r] & USp(2N-2) \ar@{-}[r] &SO(2N-1) \ar@{-}[r] 
		 \ar@{-}[r] & \dots \ar@{-}[r]
		 & SO(5) \ar@{-}[r] &USp(2)  \ar@{-}[r] & SO(3)	}
\end{tabular}
\end{center}
so we can compute the central charges using the field content
\ie
n_v={N(4N^2+3N+2)\over 3},\quad 
n_h={N(4N^2+6N-1)\over 3}
\fe
which gives
\ie
a={N(8N^2+7N+3)\over 24},\quad 
c={N(1+2N)^2\over 12}.
\fe
Moreover the flavor central charge is supplied by $2N+1$ half hypers in the fundamental representation of $USp(2N)$ thus
\ie
k=N+{1\over 2}
\fe
Everything above is consistent with \eqref{kaceg4}.
 
\end{eg}

\begin{eg}
  Let's take $G=B_N$ which is derived from the $\mZ_2$ twist of $A_{2N-1}$ $(2,0)$ theory. The twisted irregular puncture is specified by
    \begin{equation}
  b_t=2N,~~k=-N
  \end{equation}
  and then $\Delta[z]=2$. The Coulomb branch spectrum and central charges from our general formulae
   \ie
    &\Delta=\bigsqcup_{j=1}^{N-1}\{2i|1\leq i\leq j\} \bigsqcup_{j=1}^{N-1}\{2i|1\leq i\leq j\} 
%    \Delta=&\{2N-2,~2N-4,~\dots,2
%\\
%&2N-2,~2N-4,~\dots,2,
%\\
%&\dots
%\\
%&2
%\\
%&2
%\}\\
\\
&k_{SO(2N+1)}=2N-2,
\quad
a={N(8N^2-5N-3)\over 24},\quad 
c={N(2N^2-N-1)\over 6}.
\label{kaceg5}
\fe
From Table~\ref{table:aodata} the irregular singularity contributes no additional mass parameters and $2N-2$ marginal couplings.

 The theory has a Lagrangian description as
\begin{center}
\begin{tabular}{ccc}
%\xymatrixrowsep{.1in}
\xymatrixcolsep{.1in}
	\xymatrix{
		\boxed{SO(2N+1)}\ar@{-}[r] & USp(2N-2) \ar@{-}[r] &SO(2N-1) \ar@{-}[r] 
		 \ar@{-}[r] & \dots \ar@{-}[r]
		 & SO(5) \ar@{-}[r] &USp(2)  \ar@{-}[r] & SO(3)	}
\end{tabular}
\end{center}
so we can compute the central charges using the field content
\ie
n_v={N(4N^2-3N-1)\over 3},\quad 
n_h={4N( N^2 -1)\over 3}
\fe
Hence
\ie
a={N(8N^2-5N-3)\over 24},\quad 
c={N(2N^2-N-1)\over 6}.
\fe
Moreover the flavor central charge is supplied by $2N-2$ half-hypers in the fundamental representation of $SO(2n+1)$ thus
\ie
k_{SO(2N+1)}=2N-2
\fe
Again we see they are in agreement with \eqref{kaceg5},
	\end{eg}

\begin{eg}
 Half-hypermultiplet 
   
A free half-hyper multiplet in the fundamental representation of $USp(2N)$ flavor symmetry can be constructed using $A_{2N}$ $(2,0)$ theory with $\mZ_2$ twist. The irregular puncture is specified by
\ie
  b_t=4N+2,~~k=1-(4N+2)
\fe
It is easy to see from our general formulae that the Coulomb branch is empty in this case and the central charges are
\ie
k_{USp(2N)}={1\over 2},~~2a=c={N\over 12}
\fe
as expected for a half-hyper in the fundamental representation of $USp(2N)$ (or $N$ free half-hypers).

We emphasize here that this is the only twisted theory within our construction that has an empty Coulomb branch yet nonvanishing central charge.
\label{eg:hh}
\end{eg}

 \begin{eg}
    Let's take $G=C^\text{\sout{anom}}_{N-1}$ which is derived from the $\mZ_2$ twist of $D_{N}$ $(2,0)$ theory. We take $N=3n$ with $n\in \mZ^+$ and the irregular puncture is specified by
    \ie
      b_t=6n,~~k=3-6n
\fe
  so that $\Delta[z]=2n$.
   The Coulomb branch spectrum and central charges from our general formulae
   \ie
    &\Delta=\{2,4,\dots,2n\}\bigsqcup \{2,4,\dots,4n-2\}
\\
&k_{USp(6n-2)}=2n,
\quad
a= \frac{1}{24} (66 n^2-33 n+5 )
,\quad 
c=3 n^2-\frac{3 n}{2}+\frac{1}{6}
\label{kaceg7}
\fe
From Table~\ref{table:ddata} the irregular singularity contributes no additional mass parameters and $2$ marginal couplings.

The theory has a Lagrangian description by
 \begin{center}
	\begin{tabular}{ccc}
		\xymatrix{
			USp(2n-2) \ar@{-}[r] & SO(4n) \ar@{-}[r] & \boxed{USp(6n-2)}
		}
	\end{tabular}
\end{center}
We can then check the central charge by counting the multiplets
\ie
n_v=(n-1)(2n-1)+2n(4n-1),\quad n_h=2(n-1)\cdot 2n+2n\cdot 2(3n-1)
\fe
which gives
\ie
a= \frac{1}{24} (66 n^2-33 n+5 )
,\quad 
c=3 n^2-\frac{3 n}{2}+\frac{1}{6}.
\fe
Furthermore  the flavor central charge for $USP(6N-2)$ is
\ie
k_G=2N
\fe
from the quiver, in perfect agreement with \eqref{kaceg7}.

The simplest example in this sequence of theories is when $n=1$ which can be equivalently described by a cyclic quiver with two $SU(2)$ nodes, 
\begin{center}
	\begin{tabular}{ccc}
		\xymatrix{
			SU(2) \ar@{=}[r] & SU(2)	}
	\end{tabular}
\end{center}
and comes from type $A_1$ $(2,0)$ theory on $T^2$ with two punctures. This theory has $USp(4)$ enhanced flavor symmetry which is manifest in our description from type $D_3$ $(2,0)$ theory on $\mP^1$ with twisted irregular punctures.\footnote{This theory also appears in the sequence considered in Example~\ref{eg:3} at $N=1$.}

\end{eg}

\begin{eg}
  Let's take $G=C_N^\text{\sout{anom}}$ which is engineered from the $\mZ_2$ twist of $D_{N+1}$ $(2,0)$ theory. We take $N=3n$ and the irregular singularity specified by
\ie
  b_t=6n,~~k=3-6n
\fe
  so that $\Delta[z]=2n$.    The Coulomb branch spectrum and central charges from our general formulae
   \ie
    &\Delta=\{2,4,\dots,2n\}\bigsqcup \{2,4,\dots,4n\}\bigsqcup \{2n+1\}
\\
&k_{USp(6n)}=2n+1,
\quad
a= \frac{1}{24} (66 n^2+45 n+5 )
,\quad 
c=3 n^2+\frac{3 n}{2}+\frac{1}{6}
\label{kaceg8}
\fe
From Table~\ref{table:ddata} the irregular singularity contributes one mass parameter and $2$ marginal couplings.
 
 The Lagrangian description is given by
\begin{center}
	\begin{tabular}{ccc}
		\xymatrix{
			\boxed{USp(6n)}  \ar@{-}[r] &SO(4n+2) \ar@{-}[r] & USp(2n) \ar@{-}[r] & \boxed{SO(2)}
		}
	\end{tabular}
\end{center}
From the number of multiplets
\ie
n_v=(2n+1)(4n+1)+n(2n+1),\quad n_h=3n\cdot (4n+2)+(4n+2)\cdot n+2n,
\fe
we obtain the conformal central charges from \eqref{weakcc} 
\ie
a=  \frac{1}{24} \left(66 n^2+45 n+5\right)
,\quad 
c= 3 n^2+2 n+\frac{1}{6},
\fe
as well as the flavor central charge
\ie
k_{USp(6n)}=2n+1,
\fe
in agreement with \eqref{kaceg8}.

 \end{eg}

 \begin{eg}

Consider the $F_4$ theory constructed from $E_6$ $(2,0)$ theory with $\mZ_2$ twist and the irregular puncture is specified by
\ie
b_t=12,~k=-9.
\fe
so that $\Delta[z]=2$. The Coulomb branch spectrum and central charges can be read off from our general formulae
\ie
&\Delta=\{10,8,6,4,2,8,6,4,2,6,4,2,4,2,4,2\}  
\\
&k_{F_4}=8,~~a={203\over 6},~~c={104\over 3}.
\label{kaceg9}
\fe
From Table~\ref{table:e6data} the irregular singularity contributes no mass parameter and $5$ marginal couplings.

The theory has a Lagrangian description by 
\begin{center}
	\begin{tabular}{ccc}
		\xymatrix{
			&   & \boxed{SO(1)} \ar@{-}[d] &   &  
			\\
			  &   &USp(4)\ar@{-}[d] &   &   
			\\
			\boxed{SO(9)}  \ar@{-}[r] &USp(8) \ar@{-}[r]  &SO(11)\ar@{-}[r] & USp(6) \ar@{-}[r] &  {SO(5)}
		}
	\end{tabular}
\end{center}
By counting multiplets
we obtain the conformal central charges
\ie
a={203\over 3},~~c={104\over 3},
\fe
as well as the flavor central charge for the $SO(9) \subset F_4$
\ie
k_{SO(9)}=8.
\fe
They are consistent with \eqref{kaceg9} since the index of embedding $I_{\mf{so}_9\hookrightarrow \mff_4 }=1$	and our description with irregular puncture makes manifest the enhanced $F_4$ flavor symmetry of the theory.\footnote{Recall from \cite{Argyres:2007cn} that the Dynkin index of embedding for $G\subset J$ is computed by 
 	\ie
 	I_{G\hookrightarrow J}={\sum_i T({\bf r}_i)\over T({\bf r})}
 	\fe	
 	where ${\bf r}$ denotes a representation of $J$  which decomposes into $\oplus_i {\bf r}_i$ under $G$, and $T(\cdot)$ computes the quadratic index of the representation (which can be found for example in \cite{Yamatsu:2015npn}).
 }
 \end{eg}

\subsection{Twisted theories and non-Lagrangian conformal matter }
In addition to the Lagrangian examples discussed in the last section, our twisted theories also generalize many non-Lagrangian conformal matter theories constructed in class S with regular (tame) punctures. Below we provide various examples and their reincarnation in our construction with twisted irregular punctures. Various physical data of these conformal matter theories have been extracted from the SW geometry, superconformal index, and decoupling limits of certain Lagrangian theories. We will view these information as support for our construction of the much larger class of theories and nontrivial evidence for our conjectured formulae for the central charges \eqref{fla} and \eqref{conjcc}. As we will see in the examples, our construction often makes manifest the enhanced global symmetry which is obscure in the ordinary (regular punctures) class S setup.

\begin{eg} $R_{2,2N}$ conformal matter
	
The $R_{2,2N}$ non-Lagrangian theory was constructed in \cite{Chacaltana:2014nya} from the $\mZ_2$ twist of type $A_{2N}$ $(2,0)$ theory with three regular punctures on a sphere: one minimal untwisted puncture, and two maximal twisted punctures.\footnote{The theory was also constructed by the circle compactification of a 5d $\cN=1$ SCFT with $\mZ_2$ twist \cite{Zafrir:2016wkk}.} This theory also arises in the decoupling limit of $\cN=2$ $SU(2N+1)$ SYM coupled to one symmetric and one antisymmetric rank two tensor hypermultiplets in a S-dual frame \cite{Chacaltana:2014nya}.

The Coulomb branch spectrum of $R_{2,2N}$ SCFT to be
\ie
\Delta=\{3,5,7,\dots, 2N+1\}
\label{R22Ncoulomb}
\fe
and the conformal central charges are
\ie
a={14N^2+19N+1\over 24},\quad c={8N^2+10N+1\over 12}.
\label{R22Nac}
\fe
The theory has enhanced $U(1)\times USp(4N)$ flavor symmetry where only the maximal subgroup $U(1)\times USp(2N)\times USp(2N)$ is manifest from the regular punctures  in $A_{2N}$. The $USp(4N)$ factor has  central charge 
\ie
k_{USp(4N)}=N+1.
\label{R22Nfc}
\fe 

Alternatively, the $R_{2,2N}$ theories can be constructed from type $A_{4N}$ $(2,0)$ theory with $\mZ_2$ twist and the twisted irregular puncture is specified by
\ie
b_t=4N,~~k=1-4N
\fe
 such that $\Delta[z]=4N$. One can immediately read off the Coulomb branch spectrum from Table~\ref{table:coulomb} and the result coincides with \eqref{R22Ncoulomb}. The manifest $USp(4N)$ flavor symmetry comes from the regular twisted puncture and its flavor central charge is determined by \eqref{fla} to be \eqref{R22Nfc}. Our description also makes obvious Witten's global anomaly for $USp(4N)$ \cite{Tachikawa:2018rgw}. From Table~\ref{table:aedata} we see that the irregular puncture provides the additional mass parameter responsible for the  $U(1)$ factor in the flavor symmetry. It is also easy to check that the central charges computed from \eqref{conjcc} is consistent with the result \eqref{R22Nac} from \cite{Chacaltana:2014nya}.
  
\end{eg}

\begin{eg}
 {$USp(2N)$ conformal matter}
 
Let is consider $A_{2N}$ $(2,0)$ theory with $\mZ_2$ twist and the irregular puncture is specified by
\ie
b_t=2N,~k=1-2N
\fe
which gives $\Delta[z]=2N$. For $N$ even this is just the $R_{2,2N}$ theories in the last example. Here we focus on $N$ odd which has $USp(2N)$ flavor symmetry. Our general prescription for the twisted theories give
\ie
&\Delta=\{N+1, N-1,\dots, 2\}
\\
&k_{USp(2N)}={N\over 2}+1,~~a={(7N+5)(N+2)\over 48},~~c={(2N+1)(N+2)\over 12}.
\label{USPcmkac}
\fe
From  Table~\ref{table:aedata} we see the irregular puncture contributes one marginal coupling but no mass parameters.

For $N=5$, this is related to the $SU(2)_4\times USp(10)_{7\over 2}$ SCFT of \cite{Chacaltana:2013oka} with $\Delta=\{4,6\}$ and $(n_h,n_v)=(35,18)$ where the $SU(2)_4$ flavor symmetry is gauged by an $SU(2)$ vector multiplet.
\begin{center}
	\begin{tabular}{ccc}
		\xymatrix{
		  SU(2)  \ar@{-}[r] &	\boxed{SU(2)_4\times USp(10)_{7/2}~\text{SCFT}} 
		}
	\end{tabular}
\end{center}
For $N=3$, this is the rank one $SU(2)_4\times USp(6)_{5\over 2}$ SCFT of \cite{Chacaltana:2013oka} with $\Delta=\{4 \}$ and $(n_h,n_v)=(15,7)$  where the $SU(2)$ flavor symmetry is gauged by a $SU(2)$ vector multiplet.
 	
For $N=1$, we get the familiar $\cN=4$ $SU(2)$ SYM with $n_h=n_v=3$\,! The $SU(2)$ flavor symmetry with central charge $k_{SU(2)}={3\over 2}$ is now realized manifestly by the twisted regular puncture in $A_2$ $(2,0)$ theory. It also carries Witten's global anomaly for $SU(2)$. 
 \end{eg}

  \begin{eg}
  	{$SO(2N+1)$ and $SO(2N+1)\times U(1)$ conformal matter }
  	
  	Let's consider the $A_{2N-1}$ $(2,0)$ theory with the $\mZ_2$ twist and the irregular singularity is defined by
  	\ie
  	b_t=2N,~k=1-2N
  	\fe
  	so that $\Delta[z]=2N$. From the general formulae, we have
  	\ie
  	&\Delta=\{N-1-2i>1|i\geq 0\}
  	\\
  	&k_{SO(2N+1)}=N-1
  	\fe 
  	and the conformal central charges
  	\ie
  	N~{\rm even}:~&a={7N^2-5N-10\over 48},~~c={2N^2-N-2\over 12},
  	\\
  	N~{\rm odd}:~&a={7N^2-5N-2\over 48},~~c={2N^2-N-1\over 12}.
  	\fe
  	From Table~\ref{table:aodata}, in the $N$ even case  the irregular puncture contributes no marginal couplings but one additional mass parameter, whereas the $N$ odd case has one marginal coupling and no extra mass parameter.  
 
Upon closer inspection, it turns out that this $\mZ_2$ twisted class $A_{2N-1}$ setup does not make manifest the full flavor symmetry. For $N$ odd, the theory is identical to the $USp(N-1)$ SYM conformally coupled to $N+1$ fundamental flavors which has $SO(2N+2)$ symmetry. For $N$ even the theory is identified with the $R_{2,N-1}$ theories in \cite{Chacaltana:2010ks}. Our formulae above again give the correct conformal and current central charges as computed previously with standard methods.
  	
  \end{eg}

\begin{eg} {$G_2$ conformal matter}
	
Let's consider the $D_4$ $(2,0)$ theory with $\mZ_3$ twist and choose the irregular puncture to be given by
\ie
b_t=12,~k=-8.
\fe
Then $\Delta [z]=3$. Our general prescription gives
\ie
&\Delta=\{2,3,3\},~~k_{G_2}=3,~~a={27\over 8},~~c={7\over 2}.
\label{G2cm}
\fe
From  Table~\ref{table:d4data} we see the irregular puncture contributes one marginal coupling but no mass parameters.

This is identified with the $E_6$ Minahan-Nemeschansky (MN) theory \cite{Minahan:1996cj} with an $SU(3)$ subgroup of $E_6$ flavor symmetry gauged. 
\begin{center}
	\begin{tabular}{ccc}
		\xymatrix{
			SU(3)  \ar@{-}[r] &	\boxed{ (E_6)_3~\text{MN SCFT}} 
		}
	\end{tabular}
\end{center}
More explicitly, the relevant maximal Lie algebra embedding is $\mfe_6 \supset \mf{su}_3\oplus \mfg_2$ with embedding indices determined by the branching rule $\bf 27\to (6,1)\oplus (3,7)$ to be
\ie
I_{\mfg_2\hookrightarrow \mfe_6 }=1,\quad I_{\mf{su}_3\hookrightarrow \mfe_6 }=2.
\fe
Therefore the gauged $SU(3)$ global symmetry has central charge $k_{SU(3)}=6$ which ensures conformal invariance. The commutant $G_2$ becomes the residue global symmetry with  $k_{G_2}=3$ consistent with \eqref{G2cm}.

\end{eg}

\begin{eg}{$F_4\times U(1)$ conformal matter}

Let's consider the $E_6$ $(2,0)$ theory with $\mZ_2$ twist and the irregular singularity is defined by
 \ie
 b_t=8,~k=-7
 \fe
 such that $\Delta[z]=8$. The physical data from our general formulae
 \ie
 &\Delta=\{4,5\} 
 \\
 &k_{F_4}=5,~~ a={14\over 3},~~c={16\over 3} 
 \fe
The full global symmetry of the theory is $U(1)\times (F_4)_5$ where extra $U(1)$ comes from the irregular puncture (see Table~\ref{table:e6data}). 

This theory is identified with the $SO(9)_{5}\times U(1)$ SCFT in \cite{Chacaltana:2015bna}. Note that our description predicts the enhancement of flavor symmetry from $SO(9)_5$ to $(F_4)_5$.

\end{eg}

 \begin{eg}
{$F_4$ conformal matter}

Let's consider another irregular defect in the $E_6$ theory with $\mZ_2$ twist specified by
\ie
b_t=12,~k=-10.
\fe
Then $\Delta[z]=6$ and the physical data from our general formulae
\ie
&\Delta=\{2,2,6,6\}
\\
&k_{F_4}=6,~~ a={47\over 6},~~c={26\over 3} 
\fe 
From Table~\ref{table:e6data}, the irregular puncture contributes two marginal coupling but no mass parameters. 

This theory is identified with the $(E_8)_6$ MN theory and $(D_4)_2$ SW theory\footnote{This is the familiar Seiberg-Witten theory constructed by $SU(2)$ $\cN=2$ SYM coupled to 4 fundamental hypermultiplets \cite{Seiberg:1994aj}. The theory has $SO(8)$ flavor symmetry with flavor central charge $k_{D_4}=2$.} with the diagonal $(G_2)_8$ flavor symmetry subgroup gauged 
\begin{center}
	\begin{tabular}{ccc}
		\xymatrix{
		\boxed{ (D_4)_2~\text{SW SCFT}} 	\ar@{-}[r] &	G_2  \ar@{-}[r] &	\boxed{ (E_8)_6~\text{MN SCFT}} 
		}
	\end{tabular}
\end{center}
The relevant  subalgebras are $\mfg_2 \oplus \mff_4 \subset \mf{e}_8$ and $\mfg_2 \subset \mf{so}_7 \subset \mf{so}_8$. It follows from the branching rules $\bf 248\to (14,1 )\oplus(7,26)\oplus(1,52)$ and $\bf 8 \to 7\oplus 1$ that
\ie
I_{\mfg_2\hookrightarrow \mfe_8 }= I_{\mfg_2\hookrightarrow \mf{so}_8}=1.
\fe
Thus the $G_2$ diagonal subgroup of has central charge $6+2=8$ which ensures conformal invariance.

	\end{eg} 

%%%%%%%%%%%%%%%%%%%%%%%%%%%%%%%%%

\section{Vertex operator algebra of twisted theories}
\label{sec:voa}

It was shown in \cite{Beem:2013sza} that for any 4d
$\mathcal{N}=2$ SCFT, one can associate a 2d chiral algebra or vertex operator algebra (VOA). The basic correspondence is as follows:

\begin{itemize}
\item The 2d Virasoro central charge $c_{2d}$ is given in terms of the conformal anomaly $c_{4d}$ of the 4d theory as
\ie
	c_{2d} = -12 c_{4d}\,. 
\fe
\item The global symmetry algebra $\mathfrak{g}$ becomes an affine Kac-Moody algebra $\hat{\mathfrak{g}}_{k_{2d}}$ and the level of affine Kac-Moody algebra $k_{2d}$ is related to the 4d the flavor central charge $k_F$ by
\begin{equation}
	k_{2d}=- k_F \ . 
\label{eq:centralchargerelation}
\end{equation}
\item The (normalized) vacuum character of the chiral algebra/VOA is identical to the Schur index of the 4d $\cN=2$ theory:
\begin{align}
 \chi_0(q) = \text{I}_{\text{Schur}}(q) \ . 
\end{align}
\end{itemize}

We focus on the twisted theory where there is no mass parameter from the irregular singularity, and the regular singularity is labeled by a nilpotent orbit $Y$.\footnote{We use the Nahm label, which is classified by the nilpotent orbits of the flavor symmetry group $G$. Note that the Hitchin label is classified by the nilpotent orbits of Langlands dual group $G^\vee$ which is generated by the invariant subalgebra of the outer-automorphism of ADE Lie algebra (see Table~\ref{table:outm}).} Our proposal for the corresponding VOA is following

\setcounter{theorem}{2}
\begin{conj}
The VOA for the twisted theory is the W-algebra $W^{k_{2d}}(G,Y)$. Here $G$ is the flavor symmetry corresponding to the maximal twisted regular puncture, and $Y$ labels the corresponding nilpotent orbit. This W algebra is derived as the Drinfeld-Sokolov reduction of the Kac-Moody algebra $\widehat{\mfg}_{k_{2d}}$ associated to the simple Lie algebra $\mfg$ at  level $k_{2d}$.  
\end{conj}

The ADE cases of these W-algebras is considered in \cite{Xie:2016evu,Song:2017oew}, and here we discover the correspondence for non-simply-laced simple Lie groups. We have verified the relation between the 4d central charges and the 2d central charges.

\subsubsection*{Admissible levels of the 2d current algebra}
A 2d current algebra level is called admissible if it can be written in one of the following forms:
\ie
& k_{2d}=-h+{p\over q},~~(p,q)=1,~~p\geq h, ~~G=ADE,  \\
& k_{2d}=-h^{\vee}+{p\over q},~~(p,q)=(q,r^\vee)=1,~~p\geq h^{\vee},~~G=BCFG,  \\
& k_{2d}=-h^{\vee}+{p\over r^\vee q},~~(p,q)=(p,r^\vee)=1,~~p\geq h,~~G=BCFG.  
%\label{admi}
\fe

Recall the 2d levels in our case are  given by following formula: 
\begin{equation}
k_{2d}=-h^{\vee}+{1\over n}{b_t\over k+b_t}~{\rm with }~k+b_t\geq 1
\end{equation}
where  $b_t$ takes the values of $d_t$ in Table~\ref{table:dt} and $n$ is as listed in Table~\ref{table:lie}. We observe that the 2d current algebra levels of the class I twisted  theories are admissible (for generic $k'\in \mZ$): 
 \ie
& B_N~{\rm Class~ I}:k_{2d}=-h^{\vee}+{2N-1\over 2k'+1+4N-2},~~b_t=4N-2
\\
& C_N^\text{{anom}}~{\rm Class~I}:k_{2d}=-h^{\vee}+{1\over2}{2N+1\over 4N+2+2k'+1},~~b_t=4N+2
\\
& C_N^\text{\sout{anom}}~{\rm Class~ I}:k_{2d}=-h^{\vee}+{N+1\over 2N+3+2k'},~~b_t=2N+2
\\
&G_2~{\rm Class~I}:k_{2d}=-h^{\vee}+{4\over 12+3k'\pm1},~~b_t=12
\\
&F_4~{\rm Class~I}:k_{2d}=-h^{\vee}+{9\over 18+2k'+1},~~b_t=18.
\label{admi}
\fe
Note that for these cases the 
relevant torsion outer-automorphism that defines the twisted irregular defect is always generated by the twisted Coxeter element.

The Schur index of the twisted theory is then identified with the vacuum character of vertex operator algebra, which is particularly simple for the boundary levels  \cite{kac2017remark}:
\ie
k_{2d}=-h^{\vee}+{h^{\vee}\over q},~~(h^{\vee},n)=1,~(h^{\vee},q)=1.
\fe
Comparing with the 2d levels in our list, we see that the boundary levels appear for the $B_N,~C_N^\text{\sout{anom}},~G_2,~F_4$ cases.

\subsubsection*{Associated variety and Higgs branch} 
The Higgs branch of the 4d $\cN=2$ SCFT is identified with the associated variety $X_{\cal V}$ of the VOA \cite{Song:2017oew,Beem:2017ooy}
\ie
\cM_{\rm Higgs}=X_{\cal V}.
\fe
 For affine Kac-Moody algebra with an admissible level (see \eqref{admi} which corresponds to the case $Y=F$ and we have a maximal regular puncture), the associated variety $X_{\cal V}=X_M$ is found to be the closure of certain nilpotent orbits  in \cite{arakawa2015associated}
%\begin{empheq}[box=\widefbox]{equation}
%  \bar{\nabla}^{\mu} \bar{h}_{\mu\nu} = 0
%\end{empheq}
\ie
\boxed{ 
 \begin{array}{ll}
X_M=\overline{\mathbb O}_q & {\rm for~} G=B_N,C_N^\text{\sout{anom}},G_2,F_4
\\
X_M=\overline{{}^L\mathbb O}_q & {\rm for~} G= C_N^\text{{anom}}.
\end{array}
}
\fe
We summarize the result of \cite{arakawa2015associated} for the relevant orbits $\mathbb O_q$ and ${}^L\mathbb O_q$ here in Table~\ref{table:no1}-\ref{table:no4}.  

For $q\geq h(\mfg)$, $\mathbb O_q$ is the same as the principal (maximal) nilpotent orbit $\mathbb O_{\rm prin}$ with quaternionic dimension
\ie
\dim_{\mH} \mathbb O_{\rm prin}={1\over 2}(\dim \mfg-\rank \mfg).
\fe
Similarly for $q\geq h^\vee(\mfg^\vee)$, ${}^L\mathbb O_q=\mathbb O_{\rm prin}$. 

For other values of $q$, in Table~\ref{table:no1}-\ref{table:no2}, we give the usual labelling of a nilpotent orbit of classical Lie algebras by a partition $[n_i]$. The quaternionic dimension can be easily computed by
(see \S 6 of \cite{collingwood1993nilpotent}) 
\ie
\dim_\mH \mathbb O_{[n_i]}=\begin{cases}
{N(2N+1)\over 2}-{1\over 4}\sum_i s_i^2+{1\over 4} \sum_{i\,{\rm odd}} r_i & {\rm if}~\mfg=B_N
\\
{N(2N+1)\over 2}-{1\over 4}\sum_i s_i^2-{1\over 4} \sum_{i\,{\rm odd}} r_i & {\rm if}~\mfg=C_N
\end{cases}
\fe
where $[s_i]$ is the transpose partition to $[n_i]$, and $r_j$ counts the number of appearances of the part $j$ in $[n_i]$. For exceptional Lie algebras, we use the Bala-Carter labels  \cite{collingwood1993nilpotent} for the nilpotent orbits in Table~\ref{table:no2}-\ref{table:no1}. We also include the quaternionic dimensions for the reader's convenience.

Near the lower end of the list of nilpotent orbits, we have the minimal nilpotent orbit of the smallest nonzero dimension. It corresponds to the centered one-instanton moduli space of $\mfg$.
Here the minimal nilpotent orbit of $C_N$ labelled by partition $[2,1^{2N-2}]$ shows up in Table~\ref{table:no1} at $q=1$, in which case the twisted theory is nothing but $N$ free hypermultiplets (see Example~\ref{eg:hh}). On the other hand, 
 the minimal nilpotent orbits of $B_N$, $G_2$ and $F_4$ do not appear to be Higgs branches of our twisted theories of the regular semisimple type.
These are consistent with the results of anomaly matching on the Higgs branch in \cite{Shimizu:2017kzs}.

\textbf{Example}: For illustration we consider a class I $F_4$ theory in \eqref{admi}. We take $k'=-7$, so the 2d current algebra level $k_{2d}=-9+{9\over 5}$ and we have $q=5$. Looking at Table~\ref{table:no4}, we conclude 
that the corresponding Higgs branch is the closure of nilpotent orbit with label  $F_4(a_3)$ which has quaternionic dimension 20. The Coulomb branch spectrum of this theory is $\Delta=\left\{\frac{6}{5},\frac{12}{5},\frac{16}{5},\frac{18}{5},\frac{22}{5},\frac{24}{5},\frac{36}{5},\frac{42}{5}\right\}$ and the central charges are $a={247\over 15},c={52\over 3}$.

Let's consider  the more general twisted theories where the twisted irregular puncture is as before in \eqref{admi} but the twisted regular puncture is now labeled by a general nilpotent orbit $Y$ of the group $G$.
The associated variety $X_{\cal V}$ is then given by
\ie
X_{\cal V}=X_M \cap S_Y.
\fe
which describes the Higgs branch for these theories.
Here $S_Y$ is the Slodowy slice defined by the nilpotent orbit $Y$ \cite{Song:2017oew,Beem:2017ooy}.

\begin{table}[htb]
	\begin{center}
		\begin{tabular}{|c|c|c|}
			\hline
			Lie algebra $\mfg$& $q$ odd &${}^L\mathbb{O}_q$   \\ \hline
			$C_N$ & $q\geq 2N-1$&  
						   $\mathbb O_{\rm prin}=(2N)$
						 \\
 						  & $q< 2N-1$   & $(q+1,\underbrace{q,\ldots, q}_{even}, s),~0\leq s \leq q-1, ~s~{\rm even}$ 
			\\
			~&  & $(q+1,\underbrace{q,\ldots, q}_{\rm even}, q-1,s),~0\leq s \leq q-1,~ s~~{\rm even}$ 
			
			\\ 
%			~& even & $(q,\ldots,q,s)~0\leq s \leq q-1,~ s~\rm even$ \\ 
\hline
		\end{tabular}
	\end{center}
	\caption{Nilpotent orbits ${}^L\mathbb{O}_q$ in  $C_N$ Lie algebras}
	\label{table:no1}
\end{table}

\begin{table}[htb]
\begin{center}
\begin{tabular}{|c|c|c|}
\hline
Lie algebra $\mfg$ & $q$ odd &$\mathbb O_q$   \\ \hline
$C_N$ & $q\geq 2N $  &  
 $\mathbb O_{\rm prin}=(2N)$   
\\
& $q< 2N$ & $(\underbrace{q,\ldots, q}_{\rm even}, s),~0\leq s \leq q-1,~s~\rm even$ \\
~&  & $(\underbrace{q,\ldots, q}_{\rm even}, q-1,s),~0\leq s \leq q-1,~s~\rm even$ \\ \hline
$B_N$ & $q\geq 2N$  &  
$\mathbb O_{\rm prin}=(2N+1)$  
\\
& $q<2N$   & $(\underbrace{q,\ldots, q}_{\rm even}, s),~0\leq s \leq q, ~s~\rm odd$ \\ 
~ & ~ & $(\underbrace{q,\ldots, q}_{\rm odd}, s,1),~0\leq s \leq q-1,~ s~\rm odd$ \\ 
 \hline
\end{tabular}
\end{center}
\caption{Nilpotent orbits $\mathbb{O}_q$ in $B_N$ and $C_N$ Lie algebras}
\label{table:no2}
\end{table}

\begin{table}[htb]
\begin{center}
\begin{tabular}{|c|c|c|c|}
\hline
Lie algebra $\mfg$ & $q \neq 0{\,\rm mod\,}3$  &$\mathbb{O}_q$ & $\dim_\mH$  \\ \hline
$G_2$ & $\geq 6$ & $G_2$ & 6 \\
 & $ 4,5$ & $G_2(a_1)$ & 5\\
 & $2$ & $\tilde{A}_1$ &  4\\
 & $1$ & $0$ & 0\\ \hline
\end{tabular}
\end{center}
\caption{Nilpotent orbits $\mathbb{O}_q$ in $G_2$ Lie algebra}
\label{table:no3}
\end{table}

\begin{table}[!htb]
\begin{center}
\begin{tabular}{|c|c|c|c|}
\hline
Lie algebra $\mfg$ & $q$ odd &$\mathbb{O}_q$ & $\dim_\mH$   \\ \hline
$F_4$ & $\geq 12$ & $F_4$ & 24\\
 & $ 9 ,11$ & $F_4(a_1)$ &23 \\
 & $  7$ & $F_4(a_2)$  &22\\
 & $ 5$ & $F_4(a_3)$ &20 \\ 
 & $3$ & $\tilde{A}_2+A_1$ &18 \\
 &1&0  & 0 \\ \hline
 \end{tabular}
\end{center}
\caption{Nilpotent orbits $\mathbb{O}_q$ in $F_4$ Lie algebra}
\label{table:no4}
\end{table}

%%%%%%%%%%%%%%%%%%%%%%%%%%%%%%

\newpage

 \section{Conclusion}
 \label{sec:sum}
 
 We systematically studied  irregular codimension-two defects  twisted by outer-automorphism symmetries in 6d $(2,0)$ theories. 
They engineer 4d $\mathcal{N}=2$ Argyres-Douglas (AD) SCFTs that admit  non-simply-laced flavor groups. 
 We completed the classification of twisted irregular defects of the regular semisimple type, and the result 
 was summarized in  Table~\ref{table:SW}. Together with the classification of the ADE cases in \cite{Wang:2015mra}, we have a large class of of Argyres-Douglas 
 theories with arbitrary simple flavor groups. We outlined a simple procedure to extract their Coulomb branch spectrum, central charges and in some cases, the 2d chiral algebra and Higgs branch. One can also consider the degenerations of irregular singularities and regular singularities as in \cite{Wang:2015mra,Xie:2017vaf,Xie:2017aqx} which will give rise to many new AD theories.
 
The theories we constructed here should be thought of building blocks towards a better understanding of the full space of 4d $\cN=2$ SCFTs.
On one hand, by conformally gauging the flavor symmetries of these AD theories we can form new 4d $\cN=2$ SCFTs.  
On the other hand, some of our theories admit exact marginal deformations, and it is interesting to study S-duality and weakly coupled gauge theory 
 descriptions (which may involve non-Lagrangian matters) of these theories using the method in \cite{Xie:2017vaf,Xie:2017aqx}.
 
In this large space of theories, we saw various relations between the physical data, such as those between the central charges and Coulomb branch spectrum, that call for explanations. For example it would be nice to develop field-theoretic proofs for our conjectured formulae for the central charges \eqref{conjcc} and \eqref{fla} perhaps along the lines of \cite{Shapere:2008zf}. It would also be interesting to verify that the vacuum character of our proposed VOA matches with the Schur index of the 4d theory using the 2d TQFT \cite{Gadde:2009kb,Gadde:2011ik,Song:2015wta,Buican:2017uka}.
  
Lastly, we  identified the VOA for the twisted theories defined using irregular defects which do not carry any flavor symmetry. 
It would be interesting to identify the VOA for the remaining theories.\footnote{This includes in particular the $R_{2,2N}$ theories of \cite{Chacaltana:2014nya}.}   

We hope to address some of these directions in future work.

\section*{Acknowledgments} 

We would like to thank Davide Gaiotto and Yuji Tachikawa for useful discussions and correspondences. We also thank Yuji Tachikawa for reading an earlier draft of this paper and giving many helpful comments. YW would also like to thank Yuji Tachikawa and Gabi Zafrir for collaboration on related subjects. YW is grateful for the Galileo Galilei Institute for Theoretical Physics for the hospitality during the completion of this work. The work of DX is supported by Center for Mathematical Sciences and Applications at Harvard University, and in part by the Fundamental Laws Initiative of
the Center for the Fundamental Laws of Nature, Harvard University.
 YW is supported in part by the US NSF under Grant No.~PHY-1620059 and by the Simons Foundation Grant No.~488653.

\bibliography{ref} 
\bibliographystyle{JHEP}

\end{document}